# Improving stability of moving particle semi-implicit method by source terms based on time-scale correction of particle-level impulses

**Liang-Yee Cheng**
Department of Construction Engineering, Polytechnic School of University of São Paulo
Av. Prof. Almeida Prado, trav. 2, 83 - Cidade Universitária, 05508-070, São Paulo, SP, Brazil
e-mail: cheng.yee@usp.br

**Rubens Augusto Amaro Junior (Corresponding author)**
Department of Construction Engineering, Polytechnic School of University of São Paulo
Av. Prof. Almeida Prado, trav. 2, 83 - Cidade Universitária, 05508-070, São Paulo, SP, Brazil
e-mail: rubens.amaro@usp.br

**Eric Henrique Favero**
Department of Construction Engineering, Polytechnic School of University of São Paulo
Av. Prof. Almeida Prado, trav. 2, 83 - Cidade Universitária, 05508-070, São Paulo, SP, Brazil
e-mail: eric.favero@usp.br

**Abstract:** The aim of this paper is to investigate the unstable nature of pressure computation focusing on incompressible flow modeling through the projection-based particle methods. A new approach from the original viewpoint of the momentum conservation regarding particle-level collisions is proposed to derive new source terms of pressure Poisson equation (PPE). This results in more stable computations with drastic reduction of unphysical pressure oscillations and more robust computation with pressure magnitudes almost independent to time step. Moreover, compared to other strategies, no additional computational effort is required, its implementation is extremely simple, and the only numerical parameter is the propagation speed of the perturbations, of which the calibration is much more straightforward due to its physical meaning. Simulations were carried out using moving particle semi-implicit (MPS) method improved by the proposed approach. The comparisons of computed results with theoretical and experimental ones confirmed the effectiveness of the proposed approach.

## 1. Introduction

Many numerical methods have been proposed to predict hydrodynamic loads to solve multidisciplinary problems in hydraulic, coastal, marine, and offshore areas, and grid or mesh-based methods are generally applied for such problems. However, these methods suffer from some limitations, especially for interface problems, e.g., violent free-surface flow, and numerical treatments like free surface tracking or remeshing techniques are required, resulting in a very time-consuming activity. Due to the easy implementation and flexibility to model highly complex problems, the meshless methods have opened new perspectives in recent years, especially to solve violent free-surface flows and strong fluid-structure interaction (FSI) problems. An important class of meshless method is the particle-based methods, where the domain is represented by a collection of points (particles) with its own physical properties such as mass and the internal/external forces, and of which the motions are evaluated by the interaction with the neighboring particles.

Numerical methods dealing with problems associated with strongly nonlinear interactions between water waves and structures are mainly based on solving either the fully nonlinear potential flow theory or the Navier-Stokes equations (Ma et al., 2016). In general, particle-based methods applied to free-surface flows that solved the Navier-Stokes equations can be categorized into two groups: weakly compressible and incompressible projection-based methods. The former such as smoothed particle hydrodynamics (SPH) (Gingold & Monagham, 1977; Lucy, 1977) or weakly compressible moving particle semi-implicit (WC-MPS) (Shakibaeinia & Jin, 2010) methods solve an appropriate equation of state in a fully explicit form. The latter such as moving particle semi-implicit (MPS) (Koshizuka & Oka, 1996) or incompressible smoothed particle hydrodynamics (ISPH) (Cummins & Rudman, 1999; Lo & Shao, 2002) methods solve a pressure Poisson equation (PPE) through a Helmholtz-Hodge decomposition (Bhatia et al., 2013) and application of the projection method (Harlow & Welch, 1965; Chorin, 1967; Temam, 1969). Incompressible projection-based particle methods are generally expected to provide higher accuracy in terms of hydrodynamic pressure calculation and volume conservation (Lee et al., 2008; Gotoh et al., 2013). Therefore, they are preferable in many applications focused on hydrodynamic load assessment. However, as the main challenges of these methods, compute-intensive and unstable pressure computation can be mentioned.

To achieve accurate and stable computations, substantial efforts have been carried out, as previously detailed by Khayyer and Gotoh (2016), Gotoh and Khayyer (2018) Ye et al. (2019) and Li et al. (2020). These efforts covering at least one of the following approaches:

i.) regularization for particle distribution;
ii.) high-order and/or modified discrete differential operators;
iii.) improved free-surface particle detection;
iv.) new formulations for the source term of PPE;
v.) background mesh.

In the context of a particle regularization procedure to improve non-uniform particle distribution, several methods have been proposed. Monaghan (1992) proposed the XSPH scheme which helps the particles to move with a velocity close to that of their neighboring particles, improving the smoothness of velocity field. Afterwards, Monaghan (2005) highlighted the fact that the XSPH scheme does not conserve energy and proposed an implicit XSPH to resolve this issue. Xu et al. (2009) proposed the particle shifting method, which slightly shifts the particles to prevent anisotropic particle structures, posteriorly extended by Lind et al. (2012), allowing applications to free-surface flows. Tsuruta et al. (2013) presented the so-called dynamic stabilization (DS) scheme which produces radial and anti-symmetric inter-particle forces and thus, at least, preserves both linear and angular momentum exactly. Further, Tsuruta et al. (2015) introduced a potential in void space as space potential particle (SPP) to solve inconsistency in local volume conservation and to reproduce physical motions of particles near the free surface through a particle-void interaction. In the context of the ISPH method, an optimized particle shifting (OPS) scheme, free of adjusting parameters and allowing only the tangential shifting for free-surface particles, was presented by Khayyer et al. (2017) and particle shifting technique combined to a collision model, named as hybrid particle shifting technique (HPST), was proposed by Garoosi and Shakibaeinia (2020). Khayyer et al. (2019) also applied the OPS to an improved MPS in the context of multiphase flow. The implementation process was divided into two steps, where initially the heavy phase particles are shifted ignoring light phase particles and then both heavy and light phase particles are displaced. Despite better distribution, the updating process of particles position can be computationally expensive, which is not practical (Wang & Zhang, 2018).

Instead of directly avoiding non-uniform particle distribution, modifications and corrections have been made on discrete differential operators considering irregular particle distribution. Discrete differential operators' models have been proposed to assure momentum conservation (Khayyer & Gotoh, 2008; Tanaka & Masunaga, 2010), by changing the conventional operation of subtraction in the gradient operator. The so-called higher order Laplacian model was developed by Khayyer and Gotoh (2010; 2012), by directly taking the divergence of a gradient model. Higher order approximations for differential operators were proposed based on a Taylor series expansion, giving a combination of these operators and corrective matrix (Randles & Libersky, 1996; Suzuki, 2008; Khayyer & Gotoh, 2011; Tamai & Koshizuka, 2014; Duan et al., 2018; Liu et al., 2018; Garoosi & Shakibaeinia, 2021; Jandaghian et al., 2021). However, generally these formulations are complex and demands additional computational cost to the compute-intensive methods, aside from the considerable coding implementation effort required.

Despite not being a definitive solution itself, the improvement of the free-surface particle detection techniques contributes significantly to more stable pressure computations by avoiding misdetection of free surface inside the fluid domain and the consequent mis imposition of the boundary condition for the pressure calculation. Among available fluid interface particle detection techniques, we can mention the techniques based on the sum of a weight function (Koshizuka & Oka, 1996), amount of neighborhood particles (Tanaka & Masunaga, 2010), gradient vector of a property of the particles (Itori et al., 2011), relative distance between particles (Gotoh et al., 2009), a combination of the criteria based on the weight function and the amount of neighborhood particles (Lee et al., 2011) or the weighted distribution of neighboring particles (Tsukamoto et al., 2016), threshold angles between adjacent connecting lines of a given particle and its neighbor particles (Sun et al., 2019), or a combination of purely geometric sphere covering tests based on interval analysis with an adaptive spatial subdivision (Sandim et al., 2020).

Regarding the source term of PPE, one of the considerable improvements was the combination of two error mitigating parts, the zero variation of the density and the velocity-divergence-free condition (Hirokazu, 1999; Zhang et al., 2006; Tanaka & Masunaga, 2010), which corresponds to the accumulative and instantaneous density deviations, respectively. In Hu and Adams (2007), both the zero variation of density and velocity-divergence-free constraints of the incompressibility condition were enforced through the resolution of two PPEs and via application of a fractional time-step integration algorithm. Khayyer and Gotoh (2009) and Khayyer et al. (2009) introduced a higher order source term, which allows a slight compressibility. The authors also proposed a different formulation for the calculation of density variation by the total differentiation of the weight function. Kondo and Koshizuka (2011) proposed a multi-term source consisting of one main part and two error-compensating parts. With proper selection of the coefficients multiplied by the error-compensating parts, unphysical pressure oscillation was suppressed. Khayyer and Gotoh (2011) and Gotoh et al. (2014) introduced a modified source term comprising of a set of high order error compensating, similar to the work of Kondo and Koshizuka (2011), but with dynamic coefficients as functions of the instantaneous flow field. Following the idea of dynamic coefficients, Tamai and Koshizuka (2014) and Sun et al. (2015) also derived improved source terms and demonstrated that the adoption of dynamic coefficients mitigates the non-physical pressure oscillations.. Zheng et al. (2014) proposed a new formulation of incompressible flows based on Rankine source solution. The PPE was modified into a form that does not require any direct approximations for function derivatives, then leading to a more robust numerical method, of which the main advantage is no need to approximate second-order derivatives in the PPE. Ngo-Cong et al. (2015) proposed a novel numerical approach for incompressible smoothed particle hydrodynamics by solving the PPE on a set of so-called moving integrated radial basis function networks.

Apart from aforementioned improvements, some researchers have used background meshes to mitigate problems due to the particle's maldistribution. Hwang (2011) and Ng et al. (2016) proposed an accurate MPS based on an embedded pressure mesh, namely moving particle pressure mesh (MPPM) and unstructured moving particle pressure mesh (UMPPM), to handle the continuity constraint. In both works, the main idea is to consider the pressure as Eulerian variable, which is in contrast with the original MPS whereby the pressure and velocity are computed on the Lagrangian particles. A background mesh scheme was proposed by Wang et al. (2019) aiming to an accurate, smooth and spatially continuous source term for the PPE. The source terms are first calculated at background mesh nodes with the interpolated velocities at fixed and regular neighboring nodes. Then, the computed source terms at mesh nodes are extrapolated to the nearest neighboring particles. Different from the last two works, the imaginary background meshes are only used for the calculation of the source terms of PPE, while other calculations are performed in the Lagrangian framework. The effectiveness of the approach for the enhancement of the stability and accuracy of pressure calculation is showed through several benchmark tests. The incorporation of a background mesh into Lagrangian particle-based methods seems a promising direction to solve challenging issues related to accurate and stable computations, but it may lead to complex data interpolations.

Although several efforts have been undertaken to enhance the computation of the hydrodynamic pressure, the source terms of previous works depend on the time step and, in practice, the computed pressure peak magnitudes are sensitive to time step, with an unstable behavior of increasing magnitude as the time step decreases. In summary, since the solution does not converge as time step goes to zero, these formulations are not rigorously consistent in time domain.

The important topics of consistency and stability, i.e. convergence, in particle-based methods have been addressed mainly on the spatial discretization (Souto-Iglesias et al., 2013; Ng et al., 2014; Tamai et al., 2017; Basic et al., 2018; Islam et al., 2018; Liu et al., 2019; Sigalotti et al., 2019; Macià et al., 2020). To the best knowledge of authors, only in Chen et al. (2014) the pressure oscillation associated to the temporal discretization was investigated by using the so-called NSD-MPS, in which truncated compact support are supplemented by conceptual/virtual particles with pressures solved from a modified PPE. The NSD-MPS prevents the clustering of zero-pressure free-surface particles, and as consequence the non-physical pressure oscillations could be significantly mitigated even when the time step is reduced. Within this context and aiming to provide a practical engineering tool for more time-stable assessment of loads due to extremely nonlinear hydrodynamic phenomena, an improvement on the pressure computation of the incompressible projection-based particle methods is proposed in the present work. Based on a novel viewpoint of momentum conservation in particle-level collisions, a correction of the mismatch between numerical and physical time scales is introduced to derive new formulations for the PPE. This leads to original source terms that depend directly on the spatial discretization and independent to time step, which results in time-stable computations of the pressure. Furthermore, the proposed formulation preserves advantages of the original meshless particle-based numerical methods such as their flexibility. Moreover, it is very easy in coding implementation, requires no additional computational effort on the already compute-intensive method and can be incorporated to the above mentioned approaches to improve the stability of the pressure assessment. To demonstrate the effectiveness of the new PPE source terms derived from the proposed approach, namely time-scale correction of particle-level impulses (TCPI), an in-house code based on MPS method is adopted and four representative phenomena are considered: hydrostatic condition, water jet hitting a perpendicular flat wall, dam breaking, and liquid sloshing inside a prismatic tank. The computed pressures from MPS with original and proposed source terms are compared to show the effectiveness of the proposed improvement. The numerical results are also compared against the analytical and experimental ones available in the literature or conducted by the authors. In the Appendix A, 2D standing wave simulations are conducted to demonstrate that the energy conservation features (volume and energy) are not deteriorated due the inclusion of the proposed source terms.

## 2. Numerical Method

### 2.1. Governing equations

The Lagrangian form of the conservation laws of mass and momentum for incompressible viscous flow are:

$$\frac{D\rho}{Dt} = -\rho \nabla \cdot \mathbf{u} = 0, \tag{1}$$

$$\frac{D\mathbf{u}}{Dt} = -\frac{\nabla P}{\rho} + \nu_k \nabla^2 \mathbf{u} + \mathbf{f}, \tag{2}$$

where $\rho$ is the fluid density, $\mathbf{u}$ denotes the velocity vector, $P$ represents the pressure, $\nu_k$ stands for the kinematic viscosity and $\mathbf{f}$ is the vector of the external body force per unit mass.

In the MPS method, the differential operators of the governing equations are replaced by discrete differential operators on irregular nodes derived from a positive compactly supported weight function $\omega: \mathbb{R}^{dim} \to \mathbb{R}_{\geq 0} \cup \{+\infty\}$ that accounts for the influence of a neighbor particle $j$ on a given particle. The value $dim \in \mathbb{Z}_{>0}$ denotes the spatial dimension. In this work, we adopted the Rational weight function:

$$\omega_{ij} = \begin{cases} \frac{r_e}{\|\mathbf{r}_{ij}\|} - 1 & \|\mathbf{r}_{ij}\| \leq r_e \\ 0 & \|\mathbf{r}_{ij}\| > r_e \end{cases}, \tag{3}$$

where $r_e$ is the effective radius that limits the range of the neighborhood $\Omega_i$ of the particle $i$ and $\|\mathbf{r}_{ij}\|$ represents the distance between the particles $i$ and $j$. The values of effective radius $r_e$, which define the sizes of the compact support, are chosen considering the

compromise between accuracy and computational cost. Following Koshizuka and Oka (1996), in the present work, the effective radius $r_e = 2.1l_0$ is used for the gradient operator (see Eq. (5)), divergence operator (see Eq. (6)) and particle number density (see Eq. (7)) for all simulations (2D or 3D), whereas $r_e = 4.0l_0$ and $r_e = 2.1l_0$ are used for the Laplacian operator (see Eq. (4)) in 2D and 3D simulations, respectively. $l_0$ is the initial distance between two adjacent particles.

The summation of the weight of all the neighboring particles in $\Omega_i$ is defined as the particle number density $n_i$, which is proportional to the fluid density:

$$n_i = \sum_{j \in \Omega_i} \omega_{ij}. \tag{4}$$

For an arbitrary scalar function $\phi$ and an arbitrary vector $\boldsymbol{\phi}$, the gradient, divergence and Laplacian operators are defined in Eqs. (5), (6) and (7), respectively:

$$\langle \nabla \phi \rangle_i = \frac{dim}{n^0} \sum_{j \in \Omega_i} \frac{\phi_j - \phi_i}{\|\mathbf{r}_{ij}\|^2} \mathbf{r}_{ij} \omega_{ij}, \tag{5}$$

$$\langle \nabla \cdot \boldsymbol{\phi} \rangle_i = \frac{dim}{n^0} \sum_{j \in \Omega_i} \frac{\boldsymbol{\phi}_j - \boldsymbol{\phi}_i}{\|\mathbf{r}_{ij}\|^2} \cdot \mathbf{r}_{ij} \omega_{ij}, \tag{6}$$

$$\langle \nabla^2 \phi \rangle_i = \frac{2 dim}{\lambda^0 n^0} \sum_{j \in \Omega_i} (\phi_j - \phi_i) \omega_{ij}, \tag{7}$$

where $n^0$ denotes the constant particle number density for a fully filled compact support. Finally, $\lambda^0$ is a correction parameter so that the variance increase is equal to that of the analytical solution, and is given by:

$$\lambda^0 = \frac{\sum_{j \in \Omega_i} \omega_{ij}^0 \|\mathbf{r}_{ij}^0\|^2}{\sum_{j \in \Omega_i} \omega_{ij}^0}. \tag{8}$$

To prevent particle clustering, avoiding unstable behavior when attracting forces act between particles, the pressure gradient can be calculated as (Koshizuka & Oka, 1996):

$$\langle \nabla P \rangle_i = \frac{dim}{n^0} \sum_{j \in \Omega_i} \frac{P_j - \hat{P}_i}{\|\mathbf{r}_{ij}\|^2} \mathbf{r}_{ij} \omega_{ij}, \tag{9}$$

where $\hat{P}_i = \min\limits_{j \in \Omega_i}(P_j, P_i)$ is the minimum pressure among the particle $i$ and its neighborhood. The pressure gradient of Eq. (9) can be decomposed into the following form:

$$\langle \nabla P \rangle_i = \frac{dim}{n^0} \sum_{j \in \Omega_i} \frac{P_j - P_i}{\|\mathbf{r}_{ij}\|^2} \mathbf{r}_{ij} \omega_{ij} + \frac{dim}{n^0} (P_i - \hat{P}_i) \sum_{j \in \Omega_i} \frac{\mathbf{r}_{ij}}{\|\mathbf{r}_{ij}\|^2} \omega_{ij} . \qquad (10)$$

As well discussed in the work of Khayyer and Gotoh (2013) in terms of regularity of neighboring particles and incomplete compact support, and further reported by Duan et al. (2017), by means of the reformulated Eq. (10), the first part of Eq. (10) corresponds to the original gradient model (see Eq. (5)) and the second part is a particle stabilizing term (PST), which forces particles from regions of high concentration to regions of low concentration, since $P_i - \hat{P}_i$ always give positive values. Therefore, PST is essential to avoid non-uniform distribution of particles and prevent particle clustering in MPS.

As already commented in the introduction, it is important to underline that higher order approximations were proposed for the source terms (Khayyer & Gotoh, 2009), Laplacian (Khayyer & Gotoh, 2010; Khayyer & Gotoh, 2012) and differential operators based on a Taylor series expansion (Randles & Libersky, 1996; Suzuki, 2008; Khayyer & Gotoh, 2011; Tamai & Koshizuka, 2014; Duan et al., 2018; Liu et al., 2018; Garoosi & Shakibaeinia, 2021; Jandaghian et al., 2021). However, our goal here is a critical analysis of inconsistent relation between the numerical time step and the magnitude of the pressure oscillations, with a new interpretation of the cause of the spurious pressure oscillation and providing an original method to solve it. In this way, the adoption of $0^{th}$ order operators are preferred to illustrate the improvements achieved by the proposed formulations.

## 2.2. Boundary Conditions

### 2.2.1. Free-surface boundary

In order to accurately identify the free-surface particles, the neighborhood particles centroid deviation (NCPD) technique (Tsukamoto et al., 2016) is adopted in all the simulations carried out in the present study. In the NCPD technique, a particle is defined as free-surface one and its pressure is set to zero when

$$\begin{cases} n_i < \beta_1 \cdot n^0 \\ \sigma_i > \varrho_1 \cdot l_0 \end{cases}. \quad (11)$$

The deviation $\sigma_i$ is calculated as:

$$\sigma_i = \frac{\sqrt{\left(\sum_{j\in\Omega_i}\omega_{ij}x_{ij}\right)^2 + \left(\sum_{j\in\Omega_i}\omega_{ij}y_{ij}\right)^2 + \left(\sum_{j\in\Omega_i}\omega_{ij}z_{ij}\right)^2}}{\sum_{j\in\Omega_i}\omega_{ij}}, \quad (12)$$

where $x_{ij} = (x_j - x_i)$, $y_{ij} = (y_j - y_i)$ and $z_{ij} = (z_j - z_i)$.

As recommended by Koshizuka and Oka (1996), the constant $\beta_1$ can be chosen between 0.80 and 0.99, while Tsukamoto et al. (2016) suggest $\varrho_1$ higher than 0.2. The values of $\beta_1 = 0.98$ and $\varrho_1 = 0.2$ are used for all simulations performed herein.

As the pressure on the particles identified as free-surface ones are imposed to be zero following the Dirichlet dynamic boundary condition. For the free surface particles and some inner particles where their neighbor is not enough to the correct calculation of the discrete differential operators, their pressure gradient is not properly computed. In these cases, a particle collision (PC) model is required to adjust the distances between particles, generally the free-surface ones. The collision model is applied after the explicit first stage of the MPS method, and a repulsive variation of velocity for fluid particles is enforced by:

$$\Delta \mathbf{u}_i = \begin{cases} \sum_{j\in\Omega_i} \frac{(1+\alpha_2)\mathbf{r}_{ij}\cdot\mathbf{u}_{ij}}{\alpha_3} \frac{\mathbf{r}_{ij}}{\|\mathbf{r}_{ij}\|} \frac{\mathbf{r}_{ij}}{\|\mathbf{r}_{ij}\|} & \|\mathbf{r}_{ij}\| \leq \alpha_1 l_o \text{ and } \mathbf{r}_{ij}\cdot\mathbf{u}_{ij} < 0 \\ 0 & \text{otherwise} \end{cases}, \quad (13)$$

where, according to Lee et al. (2011), values of $\alpha_1 \geq 0.8$ and $\alpha_2 \leq 0.2$ increase the spatial stability in simulations. For all cases analyzed herein, the coefficients $\alpha_1$ and $\alpha_2$ are set to 0.8 and 0.2, respectively. If the neighbor particle $j$ is fluid then $\alpha_3 = 2$, otherwise $\alpha_3 = 1$.

### 2.2.2. Rigid wall boundary

The representation of rigid wall boundary condition is done by imposing three layers of particles. The particles that form the layer in contact to the fluid are denominated wall particles ($\partial\Omega_{\text{wall}}$), of which the pressure is computed by solving PPE linear system (see Eq. (20)), together with the fluid particles. The remaining two layers are composed by dummy particles, which are used to assure the correct calculation of the particle number density of the wall particles. The nonhomogeneous Neumann boundary condition of pressure is applied at rigid walls and the pressure of a dummy particle $j$ can be approximated by:

$$P_j = P_i + \|\mathbf{r}_{ij}\| \left.\frac{\partial P}{\partial n}\right|_{\partial \Omega_{\text{wall}}}, \qquad (14)$$

with the following simplified relation (Matsunaga et al., 2020):

$$\left.\frac{\partial P}{\partial n}\right|_{\partial \Omega_{\text{wall}}} = \rho \mathbf{n}|_{\partial \Omega_{\text{wall}}} \cdot \mathbf{g} \approx \rho \frac{\mathbf{r}_{ij}}{\|\mathbf{r}_{ij}\|} \cdot \mathbf{g}. \qquad (15)$$

Eq. (14) is included in the PPE, see Eq. (20), where the second term, $\|\mathbf{r}_{ij}\| \left.\frac{\partial P}{\partial n}\right|_{\partial \Omega_{\text{wall}}}$, can be moved to the right-hand side, i.e., the source term, of the linear system.

### 2.3. Algorithm

To solve the incompressible viscous flow, a semi-implicit algorithm is used in the MPS method, which is similar to the projection method (Harlow & Welch, 1965; Chorin, 1967; Temam, 1969). At first, predictions of the fluid particle velocity and position are carried out explicitly by using viscosity and external forces terms of the momentum conservation (see Eq. (2)):

$$\mathbf{u}_i^* = \mathbf{u}_i^t + [\nu_k \langle \nabla^2 \mathbf{u} \rangle_i + \mathbf{f}_i]^t \Delta t, \qquad (16)$$

$$\mathbf{r}_i^* = \mathbf{r}_i^t + \mathbf{u}_i^* \Delta t. \qquad (17)$$

After that, the collision model is applied, following Eq. (13), and the contribution of $\Delta \mathbf{u}_i^*$ is added to the particle velocities and positions:

$$\mathbf{u}_i^{**} = \mathbf{u}_i^* + \Delta \mathbf{u}_i^*, \qquad (18)$$

$$\mathbf{r}_i^{**} = \mathbf{r}_i^* + \Delta \mathbf{u}_i^* \Delta t. \qquad (19)$$

Then the pressures of all fluid and wall particles are calculated by solving the PPE, a linear system, considering the particle number density (PND) criterion as follows (Koshizuka & Oka, 1996):

$$\langle \nabla^2 P \rangle_i^{t+\Delta t} = \frac{\rho}{\Delta t^2} \left( \frac{n_0 - n_i^{**}}{n_0} \right), \qquad (20)$$

where $n_i^{**}$ is the particle number density calculated based on the displacement of particles obtained in the explicit calculations.

A relaxation coefficient $\gamma < 1.0$ is normally introduced to the PPE:

$$\langle \nabla^2 P \rangle_i^{t+\Delta t} = \gamma \frac{\rho}{\Delta t^2} \left( \frac{n^0 - n_i^{**}}{n^0} \right). \tag{21}$$

The idea behind the adoption of relaxation or damping coefficients is to enforce the incompressibility condition in a robust way, while suppressing spurious oscillations in the discrete PPE. Besides a detailed description, Henshaw and Kreiss (1995) and Li (2020) showed that the accuracy of a numerical method can be assured by adopting a properly tuned damping coefficient, if using split-step strategy in the same manner as projection methods, the so-called split-step finite-difference scheme (Henshaw & Petersson, 2003).

In the present study, it is important to emphasize that fine-tuned value of $\gamma$, which is shown in section 3, is adopted in attempt to achieve the best results of the original formulation. Finally, velocity and position of the fluid particles are updated based on the pressure obtained by Eq. (21):

$$\mathbf{u}_i^{t+\Delta t} = \mathbf{u}_i^{**} - \frac{\Delta t}{\rho} \langle \nabla P \rangle_i^{t+\Delta t}, \tag{22}$$

$$\mathbf{r}_i^{t+\Delta t} = \mathbf{r}_i^{**} + \left( \mathbf{u}_i^{t+\Delta t} - \mathbf{u}_i^{**} \right) \Delta t. \tag{23}$$

## 3. Time-scale Correction of Particle-level Impulses (TCPI)

In explicit numerical scheme, specifically for phenomena dominated by advection effects, the propagation of perturbations is limited for numerical stability by the Courant-Friedrichs-Lewy (CFL) condition (Courant et al., 1967), which establishes the relation between the propagation velocity of the perturbations $c_s$ and the numerical parameters:

$$C_r = \frac{c_s \Delta t}{l_0} < 1.0, \tag{24}$$

where, $C_r$ is the Courant number, $\Delta t$ is the numerical time step and $l_0$ is the spatial resolution.

For time discontinuous phenomena, such as impulsive load due to a collision or impact, this condition leads to a mismatch between physical $\delta t$ and numerical time $\Delta t$ durations of the events.

In order to understand this issue, let us consider, for example, the case of successive collisions or impacts at progressive speed $c_s$ of a portion of continuum through a sequence of panels. A visualization is provided in Figure 1 adopting as an example the collision of a volume of fluid on an inclined panel. For sake of clarity, the resolution of the numerical modeling $l_0$ is adopted as the characteristic dimension. The fluid contact is first established with the point i at the instant $t_c$. Then, the volume of fluid slides down between the points i and i+1, covering the distance $l_0$ during the physical interval $\delta t = l_0/c_s$.

If the characteristic dimension of the problem is $l_0$ (Figure 1(a)), the intervals between the collisions detected at successive panels is $\delta t = l_0/c_s$ (Figure 1(b)-(d)). In addition to this, after the portion of the continuum collides one of the panels (Figure 1(b)), the contact will occur continuously and smoothly across the panel to reaches the next one (Figure 1(d)). In this way, physically, the order of the duration of the collision or impact associated to the space interval $l_0$ covered by the panel is about $\delta t$.

Now, considering $l_0$ as the minimal spatial resolution in case of the numerical modeling of a time discontinuous phenomenon (Figure 1(e)), e.g., the collision or impact between particles or grids, which are discrete representation of continuous spatial domain, it is detected at the time step $\Delta t$ associated to the instant when the event occurs (Figure 1(f)-(g)). Moreover, in the numerical computation, apart from the change of the status prior and after the time discontinuous phenomena at the time step when the event occurs, the event is also processed within the time step when the contact was detected. This is because of the discrete time and space approximation, instead of continuous contact of a particle across another particle or panel (Figure 1(b)-(d)), the next event of contact will only occur when it reaches the following checking point i+1. As a result, despite the physical interval between collisions of a particle to successive particles or grids at a progressive speed $c_s$ remains $\delta t = l_0/c_s$, numerically the duration of the collision or impact lasts about the numerical time step $\Delta t$ (Figure 1(f)-(g)).

Considering the stability criterion shown in Eq. (24), it is clear that $\Delta t < \delta t$. This means that the successive collisions or impacts that physically must last $\delta t$ are numerically shortened to $\Delta t$ when the numerical stability criterion is satisfied. Figure 1 illustrates this mismatch between physical $\delta t$ and numerical $\Delta t$ duration of collision or impact.

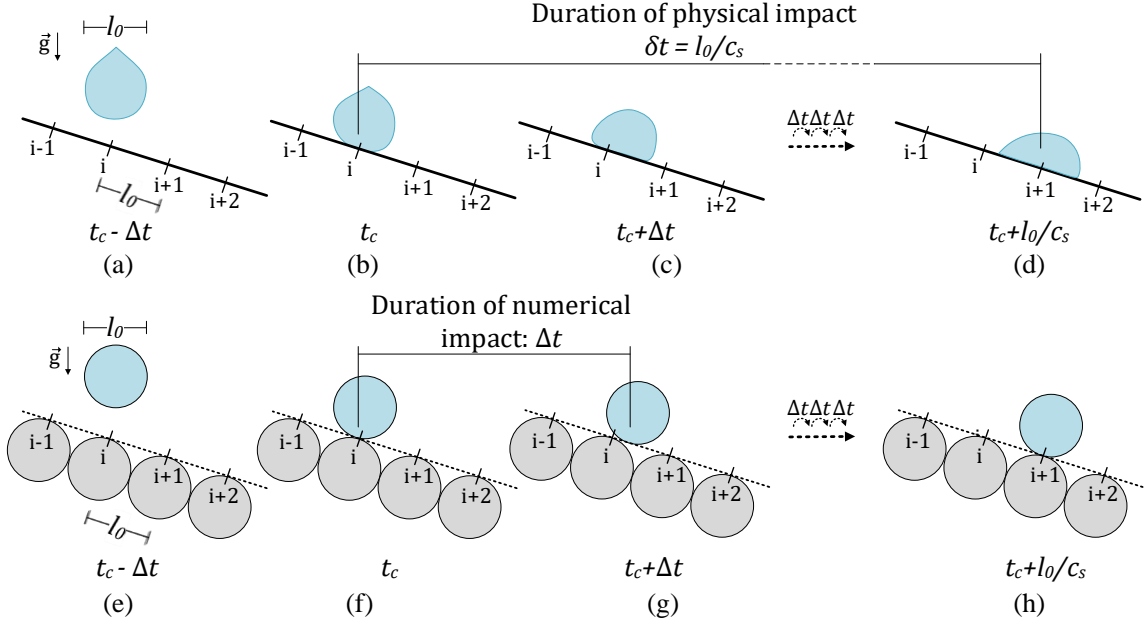

Figure 1: Continuous (top) and discrete (bottom) representations of the evolution of a volume of fluid on an inclined panel.

Since the conservation of the momentum is assured by the governing equations, the impulse **I** of a collision or impact due to loads recorded physically or computed numerically should be the same, and can be obtained by the integrating the loads considering respective time interval:

$$\mathbf{I} = \int_t^{t+\delta t} \mathbf{F}_p dt, \tag{25}$$

$$\mathbf{I} = \int_t^{t+\Delta t} \mathbf{F}_n dt, \tag{26}$$

where, $\mathbf{F}_p$ and $\mathbf{F}_n$ are collision or impact loads recorded physically and computed numerically, respectively.

Thus, the relation between $\mathbf{F}_p$ and $\mathbf{F}_n$, or the respectively pressures $P_p$ and $P_n$, becomes:

$$\frac{\mathbf{F}_p}{\mathbf{F}_n} \propto \frac{P_p}{P_n} \propto \frac{\Delta t}{\delta t} = \frac{C_r \, l_0/c_s}{l_0/c_s} = C_r \rightarrow P_n \propto \frac{P_p}{C_r} \text{ or } \Delta t \propto C_r \delta t. \tag{27}$$

The Eq. (27) shows clearly that, in the explicit schemes, the imposition of the stability criteria leads to much higher magnitude for numerical pulses than the physical ones, with the magnitude amplification coefficient of $1/C_r$. In the other words, as the duration of the numerical impulses is much shorter than the physical ones, each of these discrete impacts is accompanied by a large increase in the magnitude of the impulsive loads. The same fact was observed in Marrone et. al (2015), where the impact of two identical water jets was simulated by using an incompressible mesh-based level-set finite volume method. They observed that the computed pressure is proportional to $I_p/\Delta t$, being $I_p$ the pressure impulse, i.e., the pressure impact becomes singular with the decrease of $\Delta t$ (Meringolo et al., 2017). Moreover, a volume of fluid (VOF) technique was applied to simulate a dam break event in Mokrani and Abadie (2016) and the computed pressure peak evolves quasi linearly with respect to the inverse of $\Delta t$. Actually, instead of abruptly rise and smoothly decaying pulses recorded physically, generally a sequence of spikes with amplified peaks is computed (see, e.g., Belytschko and Mullen (1981)), as sketched in Figure 2. This is a phenomenon pointed out by Cheng and Arai (2002) as one of causes of large amplitude spurious oscillations obtained in the numerical computation of impact pressure using their mesh-based methods. It is also interesting to point out that this numerical instability is not related to the truncation error. For a rigorous study that investigate the truncation error in MPS, the reader may refer to Duan et al. (2019).

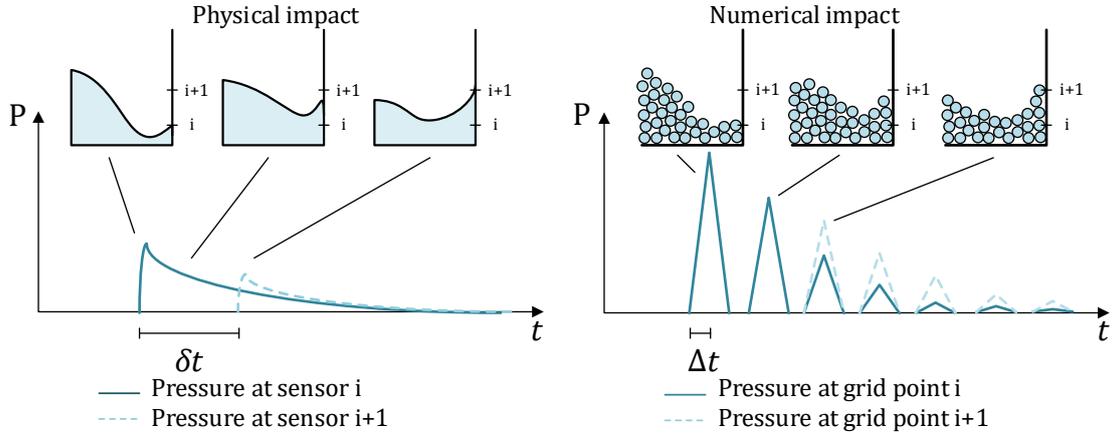

Figure 2: Sketch of physical (left) and numerical (right) impact pressure at two successive observation points. Adapted from Cheng and Arai (2002).

Besides the dilemma of the numerical stability criterion that results in the unstable computation of the time discontinuous phenomena, the solution is inconsistent in time domain, with peak values very sensitive to $\Delta t$, because the magnitude of the unstable oscillating numerical collision loads $P_n \to \infty$ when the numerical time step $\Delta t \to 0$.

As a verification of the unstable and time-inconsistent behavior of the particle-based pressure computation, an idealized scenario of a hydrostatic problem is considered. In this condition, a column of $N_P = 15$ fluid particles arranged at one-dimensional (1D) regular intervals (see Figure 3(a)) is used to illustrate that the magnitude of the computed pressure is proportional to $1/\Delta t$. Particles are governed by the equations of inviscid incompressible flow. The source term with PND deviation (see Eq. (21)) and collision model (see Eq (13)) are used here. The initial distance between particles $l_0$ is set to 0.1m, the effective radius $r_e = 2.1l_0$ is used for all operators and the relaxation coefficient $\gamma = 0.01$ is adopted. We assigned fluid density $\rho = 1\text{kg/m}^3$ and $g = 1\text{m/s}^2$. Dummy ($y = 0.0$ and $0.1$m), wall ($y = 0.2$m) and free-surface ($y = 13.0$ and $14.0$m) particles are fixed whereas the remaining inner particles are free to move. A simplified Neumann boundary condition of pressure is applied at wall and free-surface particles as follows:

$$\frac{\partial P}{\partial y} = \rho g. \tag{28}$$

It is noteworthy that, in order to verify the numerical sensibility on the computed pressures, only here we adopted the Eq. (28) for free-surface particles. Generally, the Dirichlet boundary condition of pressure is imposed at free-surface particles. However, in the present idealized 1D hydrostatic problem with fixed position of free-surface particles, this leads to a closed solution and exact pressure computation without any non-physical oscillation.

Figure 3(b) gives the computed pressure of particle initially at $y = 0.7$m for different time steps. It is clear the dependence of the magnitude of pressure fluctuations with respect to $\Delta t$. Instead of a better result that might be obtained by using a smaller $\Delta t$, high-frequency and high-magnitude pressure oscillations were computed.

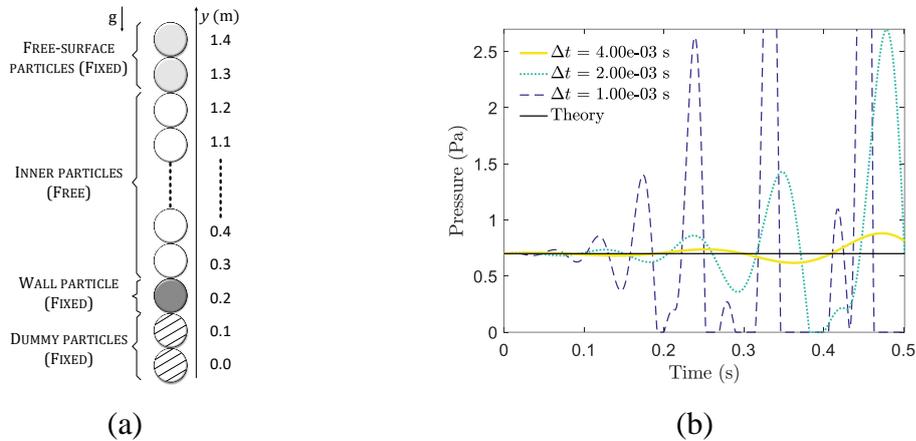

Figure 3: (a) 1D hydrostatic particle arrangement. (b) Pressure of particle initially at $y = 0.7$m. Cases simulated with $l_0 = 0.1$m and $\Delta t = 4 \times 10^{-3}$, $2 \times 10^{-3}$ and $1 \times 10^{-3}$ s.

This 1D hydrostatic example is similar to that presented in Merino-Alonso et al. (2020), where the kernel used in SPH was found to play a major role regarding spurious pressure oscillations. In this way, a numerical investigation is conducted to understand the influence of weight function on pressure oscillations. In addition to the Rational weight function given by Eq. (3), in which the weight value goes to infinite ($\omega_{ij} \to \infty$) as the distance between particles goes to zero ($\|\mathbf{r}_{ij}\| \to 0$), i.e., singular function at the position of target particle, positive non-singular weight functions $\omega: \mathbb{R}^{dim} \to \mathbb{R}_{\geq 0}$, i.e., a finite positive value $\omega_{ij} \in (0, +\infty)$ at $\|\mathbf{r}_{ij}\| = 0$, were also employed herein, namely 2$^{nd}$ order polynomial (Kondo & Koshizuka, 2011), 3$^{rd}$ order polynomial (Shakibaeinia & Jin, 2010), Wendland C2 and Wendland C4 (Wendland, 1995), given by Eqs. (29), (30), (31) and (32), respectively. The shapes of the weight functions are illustrated in Figure 4(a).

$$\omega_{2nd} = \begin{cases} \left(\dfrac{\|\mathbf{r}_{ij}\|}{r_e} - 1\right)^2 & \|\mathbf{r}_{ij}\| \leq r_e \\ 0 & \|\mathbf{r}_{ij}\| > r_e \end{cases}, \tag{29}$$

$$\omega_{3rd} = \begin{cases} \left(\dfrac{\|\mathbf{r}_{ij}\|}{r_e} - 1\right)^3 & \|\mathbf{r}_{ij}\| \leq r_e \\ 0 & \|\mathbf{r}_{ij}\| > r_e \end{cases}, \tag{30}$$

$$\omega_{C2} = \begin{cases} \left(1 - \dfrac{\|\mathbf{r}_{ij}\|}{r_e}\right)^3 \left(1 + 3\dfrac{\|\mathbf{r}_{ij}\|}{r_e}\right) & \|\mathbf{r}_{ij}\| \leq r_e \\ 0 & \|\mathbf{r}_{ij}\| > r_e \end{cases}, \tag{31}$$

$$\omega_{C4} = \begin{cases} \left(1 - \dfrac{\|\mathbf{r}_{ij}\|}{r_e}\right)^5 \left(1 + 5\dfrac{\|\mathbf{r}_{ij}\|}{r_e} + 8\dfrac{\|\mathbf{r}_{ij}\|^2}{r_e^2}\right) & \|\mathbf{r}_{ij}\| \leq r_e \\ 0 & \|\mathbf{r}_{ij}\| > r_e \end{cases}. \tag{32}$$

In order to quantify the pressure oscillations computed for the particle at $y = 0.7$m, the normalized root mean square deviation ($NRMSD_{p_{0.7}}$) was calculated until $t = 0.5$s. The $NRMSD_p$ is defined here as:

$$NRMSD_{p0.7} = \frac{RMSD_{p0.7}}{p_{y=0.7}} = \frac{1}{p_{y=0.7}} \sqrt{\frac{1}{m} \sum_{i=1}^{m} |p_{n,i} - p_{y=0.7}|^2}, \tag{33}$$

where $RMSD_{p0.7}$ denotes the root mean square deviation (also defined as the L$_2$ error norm or just L$_2$-norm), $p_n$ is the computed pressure, $p_{y=0.7}$ stands for the theoretical hydrostatical pressure of particle at $y = 0.7$m and $m$ is the number of values computed during a time interval.

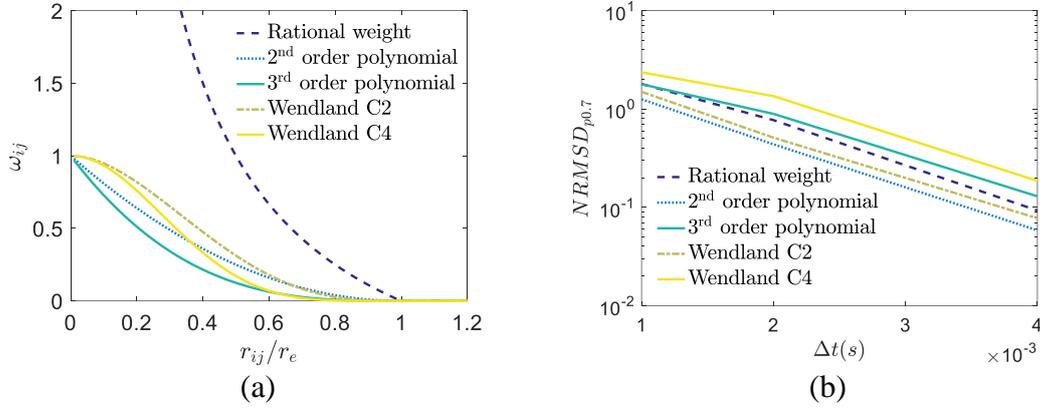

Figure 4: (a) Shape of the weight functions evaluated in present 1D hydrostatic case: Rational weight Eq. (3), 2$^{nd}$ order polynomial Eq. (29), 3$^{rd}$ order polynomial Eq. (30), Wendland C2 Eq. (31) and Wendland C4 Eq. (32). (b) Normalized root mean square deviation of pressure ($NRMSD_{p0.7}$) of particle at $y = 0.7$m as a function of the time step.

Figure 4(b) shows the logarithm of the deviations $NRMSD_{p_{0.7}}$ computed with different weight functions for three time steps $\Delta t = 4 \times 10^{-3}, 2 \times 10^{-3}$ and $1 \times 10^{-3}$ s. From Figure 4(b), the influence of the time step on the magnitude of pressure fluctuations is evident for all weight functions since the decrease of the time step leads to an increase of $NRMSD_{p_{0.7}}$, i.e., higher pressure oscillations. Moreover, lower deviations were computed by using 2$^{nd}$ order polynomial followed by Wendland C2, which reflects the small spurious pressure fluctuations compared to those computed by using the rational, 3$^{rd}$ order polynomial or Wendland C4 weight functions. Indeed, these results indicate that the non-physical pressure oscillations are affected by the choice of weight function, and some non-singular weight functions can provide more smooth and accurate pressure computation with MPS, as previously reported by Shakibaeinia and Jin (2010) and Lee et al. (2011). However, considering that the discrete differential operators of MPS are directly formulated with the weight function, one must take into account that simulations using non-singular weight functions are prone to particle clustering, resulting in the numerical simulation to diverge from the physical solution, since repulsive forces, e.g., pressure gradient term, are not remarkably increased when particles approach too close to each other. On the other hand, the singular rational weight function generally avoids particle clustering by increasing repulsive forces to infinity when two particles are too close. In this sense, the numerical stability of simulations using non-singular weight functions is highly dependent of a proper collision model or particle regularization/shifting technique in MPS. To illustrate such behavior, we simulated an hypothetical 1D hydrostatic example by using the different weight functions with and without the particle collision (PC) model given by Eq (13). Only the time step $\Delta t = 1 \times 10^{-3}$s is used. Figure 5 and Figure 6 shows the computed positions of the particles during a long time simulation of $t = 40$s using the present MPS with and without the collision model, respectively. Despite some oscillations, positions were correctly computed for all weight functions when the collision model is applied, as shown in Figure 5. On the other hand, when the collision model is not adopted, see Figure 6, only the simulation using the Rational weight function (see Eq. (3)) did not diverged, whereas particle clustering leads to unstable simulations in the remaining cases with non-singular weight functions.

Since the purpose of this study is the investigation of the relation between non-physical pressure oscillations and the time step, and given that it is independent of the weight function, see Figure 4(b), the widely-used rational weight function (Koshizuka et al., 1998; Khayyer & Gotoh, 2010; Khayyer & Gotoh, 2011; Lee et al., 2011; Tamai et al., 2017) expressed by Eq. (3), will be adopted for all the following simulations.

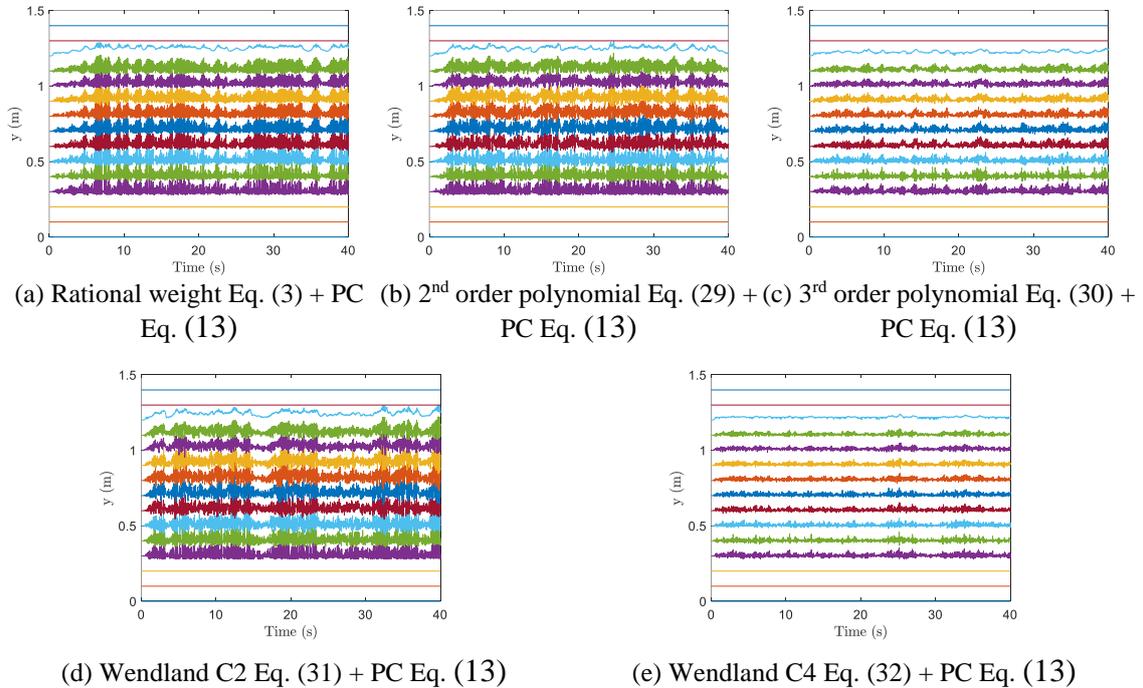

Figure 5: Particle positions in present 1D hydrostatic case during 40s computed using present MPS and considering the particle collision (PC) model given by Eq (13). Simulations using the (a) Rational weight Eq. (3), (b) 2nd order polynomial Eq. (29), (c) 3rd order polynomial Eq. (30), (d) Wendland C2 Eq. (31) and (e) Wendland C4 Eq. (32).

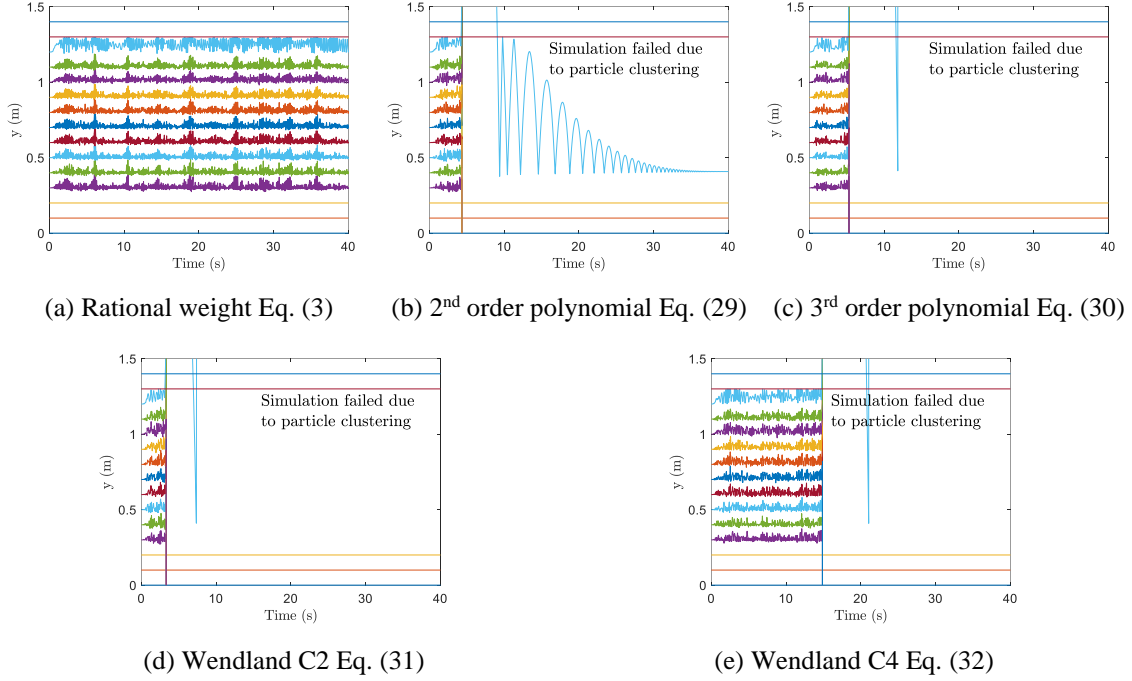

(a) Rational weight Eq. (3)  (b) 2$^{nd}$ order polynomial Eq. (29)  (c) 3$^{rd}$ order polynomial Eq. (30)

(d) Wendland C2 Eq. (31)     (e) Wendland C4 Eq. (32)

Figure 6: Particle positions in present 1D hydrostatic case during 40s computed using present MPS without the particle collision (PC) model given by Eq (13). Simulations using the (a) Rational weight Eq. (3), (b) 2$^{nd}$ order polynomial Eq. (29), (c) 3$^{rd}$ order polynomial Eq. (30), (d) Wendland C2 Eq. (31) and (e) Wendland C4 Eq. (32).

Since the essence of the spurious oscillations and the time-unstable behavior are caused by the mismatch between the numerical and physical durations of collisions or impacts, following Eq. (27), a possible solution to mitigate these issues of the numerical modeling is the application of the Courant number $C_r = \Delta t/\delta t$ as correction factor, i.e., the scale between the numerical time step and physical time interval, to compute the impulsive loads. Also, following the example shown in Figure 3, even in the hydrostatic condition spurious pressure oscillations may occur and they are clearly related to the collisions that occur in particle-level, and is an intrinsic feature of discrete Lagrangian particle-based modeling.

Now focusing on Lagrangian particle-based methods, in which the particle-level collisions occur, the correction of the mismatch between numerical and physical intervals is introduced to derive a new formulation for the PPE, by using the correction factor $C_r$ to improve the time-stability of the computed pressure. Instead of using the numerical time step $\Delta t$ in the source term (see Eq. (20)), the physical interval $\delta t$ is adopted to adjust both the magnitude and duration of the impulses, so that:

$$\langle \nabla^2 P \rangle_i^{t+\Delta t} = \frac{\rho}{\delta t^2} \left( \frac{n^0 - n_i^{**}}{n^0} \right). \tag{34}$$

This is the same as applying the correction factor to the relaxation coefficient (see Eq. (27))

$$\gamma = C_r^2 = \left( \frac{c_s \Delta t}{l_0} \right)^2 \tag{35}$$

in Eq. (21). Rewriting the PPE, we have:

$$\langle \nabla^2 P \rangle_i^{t+\Delta t} = \frac{c_s^2 \Delta t^2}{l_0^2} \frac{\rho}{\Delta t^2} \left( \frac{n^0 - n_i^{**}}{n^0} \right), \tag{36}$$

resulting in:

$$\langle \nabla^2 P \rangle_i^{t+\Delta t} = c_s^2 \frac{\rho}{l_0^2} \left( \frac{n^0 - n_i^{**}}{n^0} \right). \tag{37}$$

Observe that the last equation is independent to $\Delta t$. This is an important aspect because it suggests that for a given spatial resolution $l_0$, the computed pressure is independent of $\Delta t$, which is a strong indication that it might solve the inconsistency in time domain.

Also, instead of pressure relaxation coefficient $\gamma$ that requires empirically calibration and adjustment, in the proposed source term the only numerical parameter is the propagation speed of the perturbations $c_s$, of which the calibration is much more straightforward due to its physical meaning (see Eq. (24)) regarding the particle-level collision dynamics.

Here it is important to explain an important feature of the replacement of the numerical time step $\Delta t$ by the spatial resolution $l_0$. With the traditional $\Delta t$ based PPE source terms and fixed relaxation coefficients, the reduction of $\Delta t$ leads to larger pressure oscillations and there is no change in the spatial resolution of the numerical model. On the other hand, the proposed PPE is based on $\delta t$, which, by introducing the correction factor, leads to $l_0$ in the source term. Since the reduction of $l_0$ means the use of a higher resolution model, the mass of the particles as well as the numerical error are also reduced so that the error in the computed impulses might not increase with the reduction of $l_0$. Furthermore, as previous explained, the reduction of $l_0$ means that the physical time $\delta t = l_0/c_s$ is also smaller, and the error of the computed impulsive load, which is related to the amplitude of the oscillations and the mismatch between $\delta t$ and $\Delta t$, might be reduced.

At this point, one should remind that the reduction of $\delta t$ associated to the reduction of $l_0$ might also demand the reduction of $\Delta t$ because the relation $\delta t/\Delta t > 1.0$ must be assured due to Courant condition. In summary, using the proposed source term Eq. (37), a smaller increase of the pressure oscillation with the reduction of $l_0$ should be expected, but it would be much smaller than that due to the reduction of $\Delta t$ when using Eq. (21).

Another incompressible condition considered in the source term is that the divergence of the velocity field should be zero. As proposed by Tanaka e Matsunaga (2010), PPE can be rewritten as:

$$\langle \nabla^2 P \rangle_i^{t+\Delta t} = \gamma \frac{\rho}{\Delta t^2}\left(\frac{n^0 - n_i^t}{n^0}\right) + \frac{\rho}{\Delta t}\langle \nabla \cdot \mathbf{u} \rangle_i^{**}. \tag{38}$$

The first term of the right-hand side refers to the accumulative density deviations, i.e., the deviation of particle number density at the current instant ($n_i^t$) from that obtained for a fully filled compact support at the initial of the simulation ($n^0$). On the other hand, the second term designates the instantaneous time variation of particle density at the current instant $t$ (Gotoh et al., 2014).

Similarly to the previous procedure of introducing a relaxation coefficient $\gamma$ to reduce spurious pressure oscillation and improve the time-consistency, by using the physical time interval $\delta t$, the first term and the second term of the right side are multiplied by $C_r^2$ and $C_r$, respectively:

$$\langle \nabla^2 P \rangle_i^{t+\Delta t} = \frac{c_s^2 \Delta t^2}{l_0^2}\frac{\rho}{\Delta t^2}\left(\frac{n^0 - n_i^t}{n^0}\right) + \frac{c_s \Delta t}{l_0}\frac{\rho}{\Delta t}\langle \nabla \cdot \mathbf{u} \rangle_i^{**}, \tag{39}$$

that results in:

$$\langle \nabla^2 P \rangle_i^{t+\Delta t} = c_s^2 \frac{\rho}{l_0^2}\left(\frac{n^0 - n_i^t}{n^0}\right) + c_s \frac{\rho}{l_0}\langle \nabla \cdot \mathbf{u} \rangle_i^{**}. \tag{40}$$

Eq. (40) is also independent to time step $\Delta t$, which strongly indicates that it might be a time-consistent formulation.

The coefficients that multiply the particle number density deviation and the divergence-free condition can be expressed by:

$$\Gamma_1 = \frac{\gamma}{\Delta t^2}, \qquad \Gamma_2 = \frac{1}{\Delta t}, \qquad \Pi_1 = \frac{c_s^2}{l_0^2}, \qquad \Pi_2 = \frac{c_s}{l_0}. \tag{41}$$

Therefore, the Eqs. (38) and (40) can be rewritten as:

$$\langle \nabla^2 P \rangle_i^{t+\Delta t} = \rho \Gamma_1 \left( \frac{n^0 - n_i^t}{n^0} \right) + \rho \Gamma_2 \langle \nabla \cdot \mathbf{u} \rangle_i^{**}, \tag{42}$$

$$\langle \nabla^2 P \rangle_i^{t+\Delta t} = \rho \Pi_1 \left( \frac{n^0 - n_i^t}{n^0} \right) + \rho \Pi_2 \langle \nabla \cdot \mathbf{u} \rangle_i^{**}. \tag{43}$$

The denominations adopted hereafter for the original and improved source terms are summarized in Table 1.

Table 1: Description of the original and proposed source terms applied in this study.

| Source term | Abbreviation |
|---|---|
| Original PND deviation Eq. (21) | O-PND |
| Proposed time-scale correction of particle-level impulses PND deviation Eq. (37) | TCPI-PND |
| Original PND deviation and divergence-free condition Eq. (38) | O-PND-DF |
| Proposed time-scale correction of particle-level impulses PND deviation and divergence-free condition Eq. (40) | TCPI-PND-DF |

## 4. Calibration of the propagation speed of the perturbations $c_s$ for TCPI-PND and TCPI-PND-DF source terms

In this section, one is intended to find the range of the speed of the perturbations $c_s$ that enables stable simulations. For this purpose, the fluid volume variation and the pressure time series computed by the proposed source terms TCPI-PND and TCPI-PND-DF are analyzed considering hydrostatic condition. Figure 7 illustrates the tank dimensions of length $L_T = 30 \times l_0$, initial water level height $H_F = 20$m, and the position of the pressure sensor SH$_1$ located at the tank bottom. Even in the hydrostatic case, where particles are initially at rest, the source term of PPE is not zero for all particles so that particle motions are induced after the corrections of particle velocity and position (Kondo & Koshizuka, 2011). Furthermore, the error on total fluid volume and pressure oscillation tend to increase as the time advances. Hence, the analysis of variations on surface height and hydrostatic pressure seems an appropriate procedure to find stable coefficients $\Pi_1 = c_s^2/l_0^2$ and $\Pi_2 = c_s/l_0$.

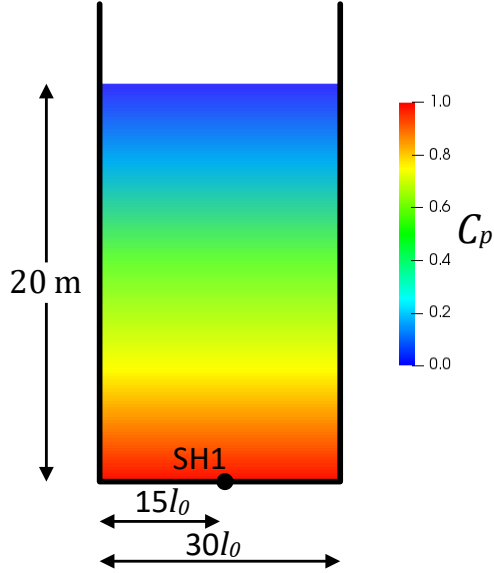

Figure 7: Calibration of the propagation speed of the perturbations $c_s$. Main dimensions of the hydrostatic tank and sensor position (SH$_1$). Color scale is representative of the pressure field in the initial configuration.

Assuming the fluid as weakly compressible, the pressure $P$ can be related to the density field $\rho$ by a linear equation of state:

$$P \sim c_s^2 \rho. \tag{44}$$

Regarding the particle-level collisions in the prediction step of the semi-implicit algorithm used in the projection-based methods, the predominant pressure scale is proportional to potential due to the gravitational effect $\rho g l_0$. Using this value in Eq. (44), the following relation is obtained:

$$c_s \sim \sqrt{g l_0}, \tag{45}$$

where $g = 9.81 \text{m/s}^2$ is the gravity.

In order to certify the above relation, several initial distances between particles $l_0 = 1.0, 0.4, 0.1$ and $0.02$m, with respective ratios $H_F/l_0 = 20, 50, 200$ and $1000$, time steps $\Delta t = l_0/20, l_0/50$ and $l_0/80$, and the propagation velocity of the perturbations ranging from $c_s = 0.5$ to $40$m/s, were adopted to simulate an hydrostatic tank during 50 seconds.

The oscillation of the dimensionless pressure $C_P = P/\rho g H_F$ is quantified by its standard deviation ($Std_P$) at sensor SH$_1$ given by:

$$Std_p = \sqrt{\frac{1}{m-1}\sum_{i=1}^{m}\left|C_{p_i} - \bar{C}_p\right|^2}, \tag{46}$$

where $m$ is the number of values computed during a time interval and $\bar{C}_p$ is the mean of the computed results $C_p$.

The variation of the fluid volume is verified by the normalized root mean square deviation of the free-surface height ($NRMSD_H$) related to its initial one $H_F$, here calculated as:

$$NRMSD_H = \frac{1}{H_F}RMSD_H = \frac{1}{H_F}\sqrt{\frac{1}{m}\sum_{i=1}^{m}|H_i - H_F|^2}, \tag{47}$$

where $RMSD_H$ denotes the root mean square deviation, $H_i$ is the computed height of the free-surface particle at the center of the tank and $m$ denotes the number of values computed during a time interval. Both deviations are computed during the last 10 seconds, i.e., the time interval between 40 and 50s.

The deviations $Std_P$ and $NRMSD_H$ of the results obtained using TCPI-PND are shown in Table 2, while those computed using TCPI-PND-DF are presented in

Table 3. In order to obtain the velocity $\tilde{c}_s$ that provides a good compromise between the minimum $Std_P$ and $NRMSD_H$ for each distance between particles $l_0$, the following weighted interpolation was adopted:

$$\tilde{c}_s = 0.5\sum_{i=1}^{n}\left[\left(\frac{\alpha_{Si}}{\sum_{i=1}^{n}\alpha_{Si}} + \frac{\alpha_{Ni}}{\sum_{i=1}^{n}\alpha_{Ni}}\right)c_{s_i}\right], \tag{48}$$

with

$$\alpha_{Si} = \frac{\max_{i\in n}|Std_{p_i}|}{Std_{p_i}} - 1 \quad \text{and} \quad \alpha_{Ni} = \frac{\max_{i\in n}|NRMSD_{H_i}|}{NRMSD_{H_i}} - 1, \tag{49}$$

where $n$ is the number of $c_s$ values considered for each initial distance $l_0$. The calculated values of $\tilde{c}_s$ are also given in Table 2 and

Table 3.

Table 2: Standard deviation of pressure ($Std_p$) and normalized root mean square deviation of the free-surface height ($NRMSD_H$) for the simulations using the proposed source term TCPI-PND. The values $c_s$ are the velocities adopted for each simulation and the calculated velocities $\tilde{c}_s$, see Eq. (48), denote optimum values that provide a good compromise between the minimum $Std_P$ and $NRMSD_H$.

| | | $l_0/\Delta t = 20\ m/s$ | | | $l_0/\Delta t = 50\ m/s$ | | | $l_0/\Delta t = 80\ m/s$ | |
|---|---|---|---|---|---|---|---|---|---|
| $l_0\ [m]$ | $c_s\ [m/s]$ | $Std_p$ | $NRMSD_H$ | $c_s\ [m/s]$ | $Std_p$ | $NRMSD_H$ | $c_s\ [m/s]$ | $Std_p$ | $NRMSD_H$ |
| 1.0 | 15 | 0.010 | 0.048 | 40 | 0.084 | 0.029 | 40 | 0.029 | 0.027 |
| | 10 | 0.005 | 0.065 | 30 | 0.012 | 0.029 | 30 | 0.013 | 0.027 |
| | 5 | 0.009 | 0.165 | 20 | 0.009 | 0.095 | 20 | 0.010 | 0.034 |
| | 1 | 0.027 | 0.227 | 10 | 0.006 | 0.047 | 10 | 0.015 | 0.041 |
| | $\tilde{c}_s = 11.20\ m/s$ | | | $\tilde{c}_s = 24.06\ m/s$ | | | $\tilde{c}_s = 26.68\ m/s$ | | |
| 0.4 | 15 | 0.024 | 0.012 | 30 | 0.055 | 0.007 | 30 | 0.086 | 0.007 |
| | 10 | 0.016 | 0.017 | 20 | 0.038 | 0.007 | 20 | 0.056 | 0.007 |
| | 5 | 0.017 | 0.023 | 15 | 0.033 | 0.009 | 15 | 0.055 | 0.008 |
| | 1 | 0.064 | 0.721 | 10 | 0.032 | 0.021 | 10 | 0.049 | 0.045 |
| | $\tilde{c}_s = 11.18\ m/s$ | | | $\tilde{c}_s = 18.30\ m/s$ | | | $\tilde{c}_s = 18.35\ m/s$ | | |
| 0.1 | 10 | 0.081 | 0.004 | 15 | 0.282 | 0.032 | 15 | 0.473 | 0.011 |
| | 7.5 | 0.072 | 0.004 | 10 | 0.339 | 0.007 | 10 | 0.289 | 0.008 |
| | 5 | 0.082 | 0.005 | 7.5 | 0.244 | 0.011 | 7.5 | 0.331 | 0.010 |
| | 1 | 0.046 | 0.021 | 5 | 0.210 | 0.396 | 5 | 0.272 | 0.086 |
| | | | | 2 | 0.283 | 0.070 | 2 | 0.079 | 0.110 |
| | $\tilde{c}_s = 4.86\ m/s$ | | | $\tilde{c}_s = 8.02\ m/s$ | | | $\tilde{c}_s = 6.96\ m/s$ | | |
| 0.02 | 5 | 0.806 | 0.003 | 5 | 0.616 | 0.009 | 5 | 0.418 | 0.017 |
| | 4 | 0.675 | 0.006 | 4 | 0.525 | 0.014 | 4 | 0.386 | 0.085 |
| | 3 | 0.541 | 0.141 | 3 | 0.437 | 0.017 | 3 | 0.262 | 0.147 |
| | 2 | 0.406 | 0.016 | 2 | 0.261 | 0.100 | 2 | 0.333 | 0.075 |
| | $\tilde{c}_s = 3.46\ m/s$ | | | $\tilde{c}_s = 3.31\ m/s$ | | | $\tilde{c}_s = 3.71\ m/s$ | | |

Table 3: Standard deviation of pressure ($Std_p$) and normalized root mean square deviation of the free-surface height ($NRMSD_H$) for the simulations using the proposed source term TCPI-PND-DF. The values $c_s$ are the velocities adopted for each simulation and the calculated velocities $\tilde{c}_s$, see Eq. (48), denote optimum values that provide a good compromise between the minimum $Std_P$ and $NRMSD_H$.

| | | $l_0/\Delta t = 20\ m/s$ | | | $l_0/\Delta t = 50\ m/s$ | | | $l_0/\Delta t = 80\ m/s$ | |
|---|---|---|---|---|---|---|---|---|---|
| $l_0\ [m]$ | $c_s\ [m/s]$ | $Std_p$ | $NRMSD_H$ | $c_s\ [m/s]$ | $Std_p$ | $NRMSD_H$ | $c_s\ [m/s]$ | $Std_p$ | $NRMSD_H$ |
| 1.0 | 15 | 0.013 | 0.042 | 40 | 0.026 | 0.029 | 40 | 0.021 | 0.023 |
| | 10 | 0.004 | 0.063 | 30 | 0.013 | 0.030 | 30 | 0.012 | 0.024 |
| | 5 | 0.004 | 0.075 | 20 | 0.007 | 0.034 | 20 | 0.009 | 0.032 |
| | 1 | 0.014 | 0.477 | 10 | 0.007 | 0.058 | 10 | 0.006 | 0.064 |
| | $\tilde{c}_s = 9.56\ m/s$ | | | $\tilde{c}_s = 24.33\ m/s$ | | | $\tilde{c}_s = 23.86\ m/s$ | | |
| 0.4 | 15 | 0.023 | 0.011 | 30 | 0.042 | 0.008 | 30 | 0.057 | 0.007 |
| | 10 | 0.016 | 0.010 | 20 | 0.030 | 0.010 | 20 | 0.035 | 0.008 |
| | 5 | 0.019 | 0.021 | 15 | 0.022 | 0.009 | 15 | 0.032 | 0.009 |
| | 1 | 0.006 | 0.115 | 10 | 0.024 | 0.009 | 10 | 0.027 | 0.011 |
| | $\tilde{c}_s = 6.80\ m/s$ | | | $\tilde{c}_s = 17.40\ m/s$ | | | $\tilde{c}_s = 18.77\ m/s$ | | |
| 0.1 | 7.5 | 0.052 | 0.004 | 10 | 0.078 | 0.006 | 10 | 0.100 | 0.008 |
| | 5 | 0.041 | 0.004 | 7.5 | 0.057 | 0.006 | 7.5 | 0.082 | 0.008 |
|      | 1   | 0.027 | 0.009 | 5   | 0.072 | 0.006 | 5   | 0.087 | 0.007 |
|      | 0.5 | 0.015 | 0.021 | 1   | 0.030 | 0.009 | 1   | 0.031 | 0.034 |
|      |     |       |       | 0.5 | 0.020 | 0.035 | 0.5 | 0.022 | 0.015 |
|      | $\tilde{c}_s = 3.30\ m/s$ | | | $\tilde{c}_s = 3.81\ m/s$ | | | $\tilde{c}_s = 3.85\ m/s$ | | |
| 0.02 | 2   | 0.341 | 0.012 | 2   | 0.325 | 0.014 | 2   | 0.302 | 0.080 |
|      | 1   | 0.243 | 0.015 | 1   | 0.031 | 0.003 | 1.5 | 0.040 | 0.004 |
|      | 0.5 | 0.010 | 0.011 | 0.5 | 0.002 | 0.055 | 1   | 0.003 | 0.039 |
|      | $\tilde{c}_s = 0.97\ m/s$ | | | $\tilde{c}_s = 0.83\ m/s$ | | | $\tilde{c}_s = 1.25\ m/s$ | | |

The computed values of $\tilde{c}_s$ and the best linear fit solution based on the least-square method, for each ratio $l_0/\Delta t = 20, 50$ and $80$ m/s, were determined for both source terms and are illustrated in Figure 8.

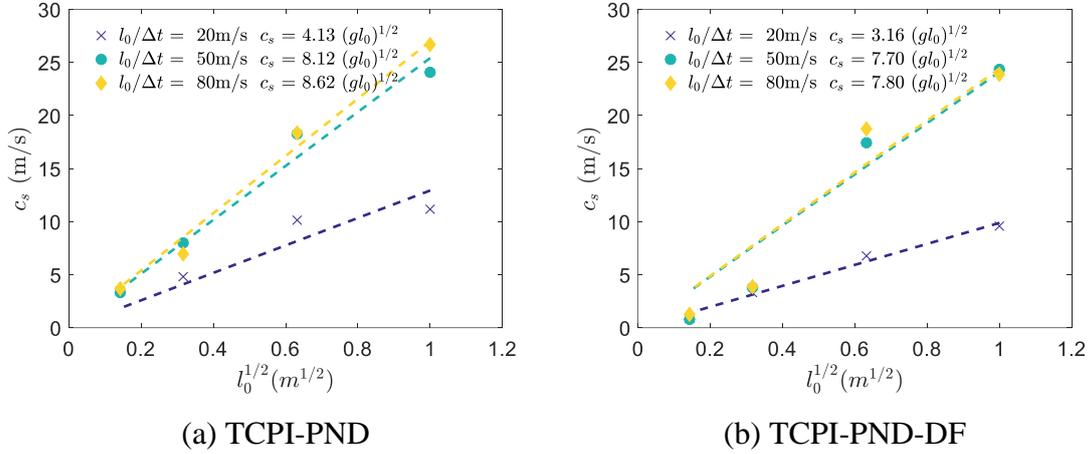

(a) TCPI-PND  (b) TCPI-PND-DF

Figure 8: Propagation speed of the perturbations $c_s$. (a) TCPI-PND and (b) TCPI-PND-DF source term.

According to Figure 8(a) and Figure 8(b), the linear relation between $c_s$ and $l_0^{1/2}$ given by Eq. (45) was numerically confirmed by the computed values of $\tilde{c}_s$ using TCPI-PND and TCPI-PND-DF, respectively. Besides the linear relation computed for all ratios $l_0/\Delta t$, lower values of $\tilde{c}_s$ were computed for the lower ratio $l_0/\Delta t = 20$m/s, while higher and very close values of $\tilde{c}_s$ were computed for the ratios $l_0/\Delta t = 50$ and $80$ m/s. This suggests that, for a given particle-based model, as smaller $\Delta t$ are used, the optimal propagation speed of perturbation $\tilde{c}_s$ converges to the same value related to $l_0$.

The computed values given in Figure 8(a) and Figure 8(b) indicate that Eq. (45) can be approximated by:

$$c_s = A\sqrt{gl_0}, \tag{50}$$

with $A \sim 10$. Indeed, considering these results and our experience, one may suggest the adoption of dimensionless constant $1 \leq A \leq 30$ to provide stable simulations by using both source terms TCPI-PND and TCPI-PND-DF. However, it is important to keep in mind that these relations were obtained by the hydrostatic pressure condition to support a more generic situation dominated by gravitational forces, e.g., water wave phenomena where inertia forces are not dominant. Hence, aiming to improve the numerical accuracy, an in-depth calibration for each simulation is recommended.

## 5. Cases of Study

In order to evaluate the performance of the source terms derived from the proposed TCPI approach, four phenomena were considered. First, a 2D hydrostatic case is simulated and the pressure of the particle on the bottom of the tank computed by the proposed source terms is compared to original ones and analytical results. After that, three simulations covering dynamic cases are considered. The first one is 2D water jet impact on a flat wall. The analytical pressure at the stagnation point and velocity on the wall are compared against the computed ones from the original and proposed source terms. The second one consisting of a 2D dam-break problem and the last one corresponds to liquid sloshing inside a 3D prismatic tank. The computed pressure at the sensor located on side wall obtained by the proposed source terms is compared to original ones and against experimental results.

The relaxation coefficient $\gamma$ needs to be calibrated based on the specificities of each case. In this sense, the definition of the value of the coefficient is carried out carefully to achieve the best results for the original formulations. As pointed out in previous works (Tanaka & Masunaga, 2010; Lee et al., 2011), a suitable relaxation coefficient range can be obtained from a hydrostatic pressure calculation. As the range $0.005 < \gamma < 0.05$ seems reasonable from the results of computed hydrostatic pressures (Tanaka & Masunaga, 2010; Lee et al., 2011), $\gamma = 0.01$ was adopted for the simulations using the original source terms. The propagation speed of the perturbations used for the simulations with the proposed source terms can be obtained by Eq. (35) as $c_s = \sqrt{\gamma} \cdot (l_0/\Delta t)$. By adopting $\gamma = 0.01$ and $l_0/\Delta t = 20$m/s, the value of $c_s = 2$m/s is obtained, which is in agreement with the recommended range given in Section 4. Since our main goal here is to demonstrate the numerical robustness and stability of the proposed source terms, we used only the value of $c_s = 2$m/s for all $l_0$, although the optimum $c_s$ is directly related to $l_0^{1/2}$, as previous detailed in section 4. The numerical parameters used for all the simulations analyzed herein are presented in Table 4. Three particle distances $l_0 = 0.010, 0.005$ and $0.002$ m and three time steps $\Delta t = l_0/20, l_0/50$ and $l_0/100$ were considered. To the authors' knowledge, the critical time steps that ensure the numerical stability in the projection particle-based methods still is an open question and generally its value is chosen according to empirical conditions based on the numerical speed of sound, the particle distance and the viscosity. The readers interested to deepen their knowledge about the choice of the maximum time from theoretical investigations in the context of incompressible particle methods are invited to read the work of Violeau and Leroy (2015).

Table 4: Numerical parameters adopted for the simulations using original (O-PND or O-PND-DF) and proposed (TCPI-PND or TCPI -PND-DF) source terms.

| $l_0$ [m] | $\Delta t$ [s] | $l_0/\Delta t$ [m/s] | $\gamma$ | $c_s$ [m/s] | $\Gamma_1[s^{-2}]$ | $\Pi_1[s^{-2}]$ | $\Gamma_2[s^{-1}]$ | $\Pi_2[s^{-1}]$ |
|---|---|---|---|---|---|---|---|---|
|  | 5.0x10⁻⁴ | 20 |  |  | 4.0x10⁴ |  | 2.0x10³ |  |
| 0.010 | 2.0x10⁻⁴ | 50 |  |  | 2.5x10⁵ | 4.0x10⁴ | 5.0x10³ | 2.0x10² |
|  | 1.0x10⁻⁴ | 100 |  |  | 1.0x10⁶ |  | 1.0x10⁴ |  |
|  | 2.5x10⁻⁴ | 20 |  |  | 1.6x10⁵ |  | 4.0x10³ |  |
| 0.005 | 1.0x10⁻⁴ | 50 | 0.01 | 2 | 1.0x10⁶ | 1.6x10⁵ | 1.0x10⁴ | 4.0x10² |
|  | 5.0x10⁻⁵ | 100 |  |  | 4.0x10⁶ |  | 2.0x10⁴ |  |
|  | 1.0x10⁻⁴ | 20 |  |  | 1.0x10⁶ |  | 1.0x10⁴ |  |
| 0.002 | 4.0x10⁻⁵ | 50 |  |  | 6.3x10⁶ | 1.0x10⁶ | 2.5x10⁴ | 1.0x10³ |
|  | 2.0x10⁻⁵ | 100 |  |  | 2.5x10⁷ |  | 5.0x10⁴ |  |

\* $\Gamma_1 = \gamma/\Delta t^2, \Gamma_2 = 1/\Delta t, \Pi_1 = c_s^2/l_0^2, \Pi_2 = c_s/l_0$.

## 6. Results and Discussions

### 6.1. **2D hydrostatic tank**

The first test case consists of a 2D hydrostatic condition with tank of length $L_T = 0.36$m and initial column water height $H_F = 0.48$m. The fluid properties are density of $\rho = 1000$kg/m³ and kinematic viscosity of $\nu_k = 1.0 \times 10^{-6}$m²/s . The computed hydrostatic pressures at the sensor SH₁, placed at the bottom of the tank, using different source terms were analyzed. The hydrostatic pressure is represented by the non-dimensional pressure coefficient $C_P = P/\rho g H_F$, and the dimensionless time ($\tau$) is defined as $\tau = t\sqrt{g/H_F}$, where $g = 9.81 \text{m/s}^2$ is the gravity acceleration.

At first, the time histories of raw pressure coefficients at sensor $SH_1$ computed using the original (O-PND) and proposed (TCPI-PND) sources terms are plotted together in Figure 9 to show the effects of the numerical parameters. It is important to point out that all the time histories shown in the present work are plot of raw results, without any filtering technique, which is a usual procedure to suppress the high frequency spurious numerical oscillation. The theoretical pressure coefficient $C_P = 1$, represented by the blue solid line, is also shown. Figure 9 shows that, when original formulation O-PND is applied, the decrease of time step, i.e., increase of the ratio $l_0/\Delta t$, leads to remarkable magnification of the magnitude of pressure oscillations, whereas for the TCPI-PND formulation proposed herein, the amplification of the pressure oscillation due to the decrease of time step is relatively small. On the other hand, Figure 9 shows that, for the TCPI-PND, despite the independence of the coefficient $\Pi_1$ in relation to the time step, for constant value $c_s = 2\text{m/s}$ adopted here, the pressure oscillations increase asymptotically as $l_0$ decreases from 0.010 to 0.002m. Also, the pressures calculated by O-PND are very sensitive to time step in all range of $l_0$, and show high oscillations for small time steps (see Figure 9($a_3$), ($b_3$) and ($c_3$)). For TCPI-PND, the influence of time step on the magnitude of pressure oscillations is much smaller than those obtained using original formulation O-PND, mainly in the cases with higher ratio $l_0/\Delta t = 100\text{m/s}$ (see Figure 9($a_3$), ($b_3$) and ($c_3$)).

Regarding the increasing pressure oscillation when $l_0$ decreases, for both original and proposed source terms, it is interesting to remind that the setting of the $c_s$ is based on the fine-tuned numerical coefficients of the original source term, which also depends on $l_0$, and $\Delta t$ (Tsukamoto et al., 2020). In this way, further improvements on the stability of the pressure computation for higher spatial resolution cases might be expected for an in-depth tuning of the parameter $c_s$ to the respective $l_0$, similar to that performed in section 4.

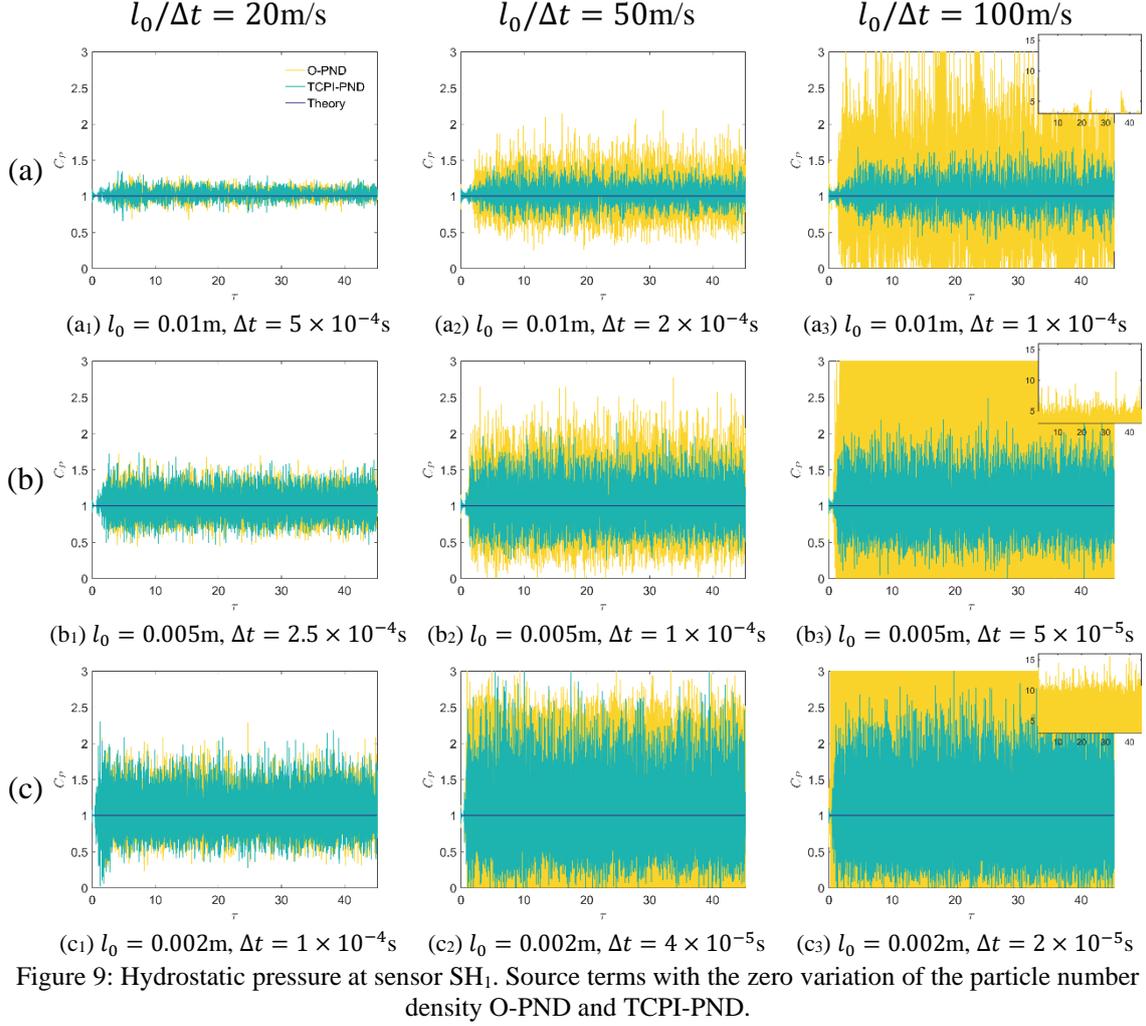

Figure 9: Hydrostatic pressure at sensor SH$_1$. Source terms with the zero variation of the particle number density O-PND and TCPI-PND.

Figure 10 shows comparisons of the time histories of the raw pressure coefficient at sensor SH$_1$ computed by the source terms O-PND-DF and TCPI-PND-DF and taken into account the numerical parameters listed in Table 4. Computed pressures and theoretical pressure coefficient $C_P = 1$ are also presented in Figure 10. Similar to the previous simulations, O-PND-DF is highly sensible to $l_0/\Delta t$ and the decrease of time step lead to very high pressure oscillations, showing the unstable issue. On the other hand, the pressures computed using TCPI-PND-DF are much more stable and independent to the $l_0/\Delta t$ or the numerical parameters. Especially for $l_0 = 0.01$ and $0.005\,\text{m}$, very small pressure oscillations are achieved for both particle distances. For $l_0 = 0.002\,\text{m}$, the results from TCPI-PND-DF shows higher oscillations but much lower than those computed by O-PND-DF. The results show remarkable improvement on the stability and the possibility of straightforward calibration provided by using $c_s$ through the application of the proposed approach, and the potential of the time-scale correction of particle-level impulses to improve the existing source terms.

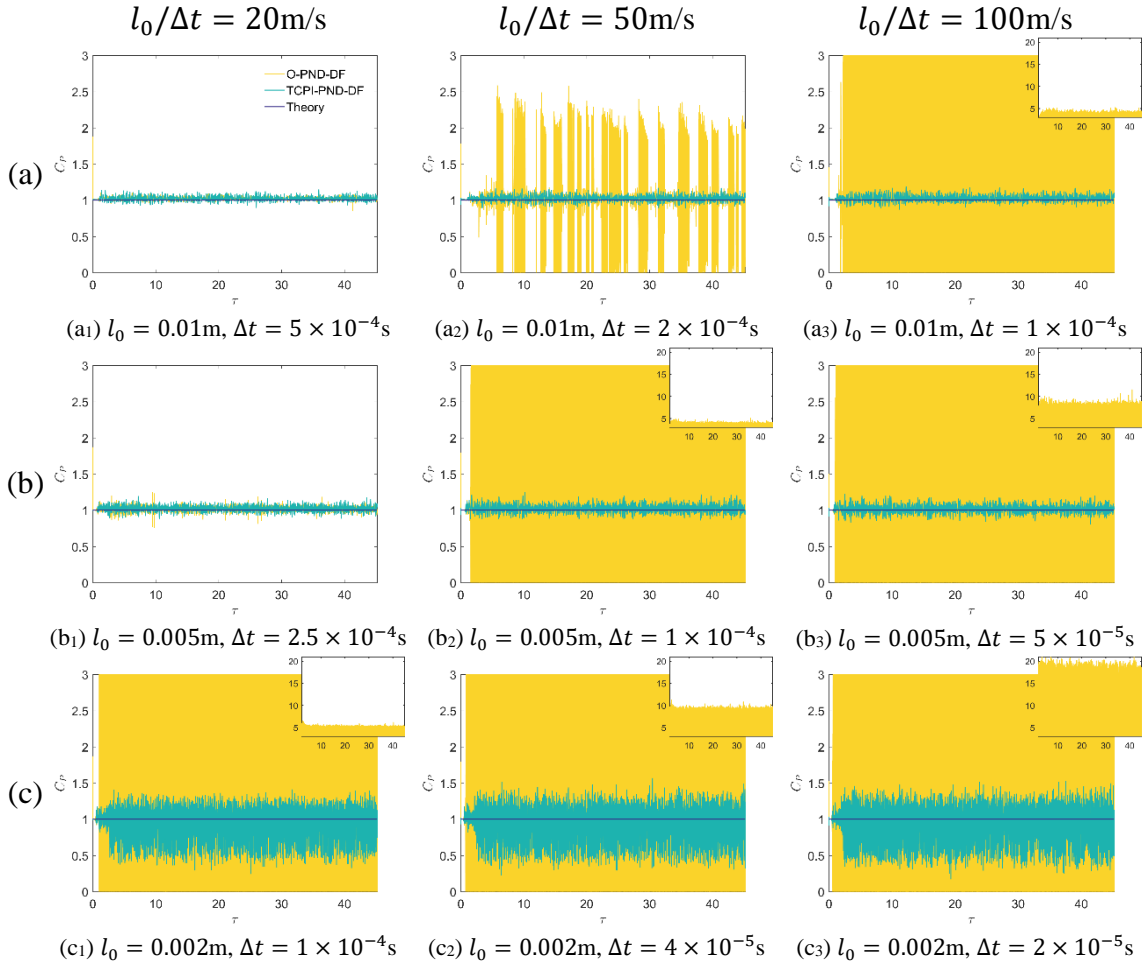

Figure 10: Hydrostatic pressure at sensor $SH_1$. Source terms with the zero variation of the particle number density and velocity-divergence-free condition O-PND-DF and TCPI-PND-DF.

Figure 11 gives the computed pressure fields for $l_0 = 0.005$m and $\Delta t = 1 \times 10^{-4}$ s (ratio $l_0/\Delta t = 50$m/s) at the instant $t = 10$s. The pressure field obtained using original source O-PND is highly irregular, which reflects the large spurious fluctuations illustrated in Figure 9(b2). The proposed TCPI-PND source term provides better pressure distribution but the computed results are not totally uniform, with some fluctuations remained, as shown in Figure 9(b2). The pressure field computed using O-PND-DF source term is more regular than that obtained by original source O-PND, but the values are inaccurate, with pressures much higher than the analytical ones, which is associated to the large pressure fluctuation shown in Figure 10 (b2). On the other hand, notable improvement is obtained by using the proposed TCPI-PND-DF source term, which gives a smooth and accurate pressure field.

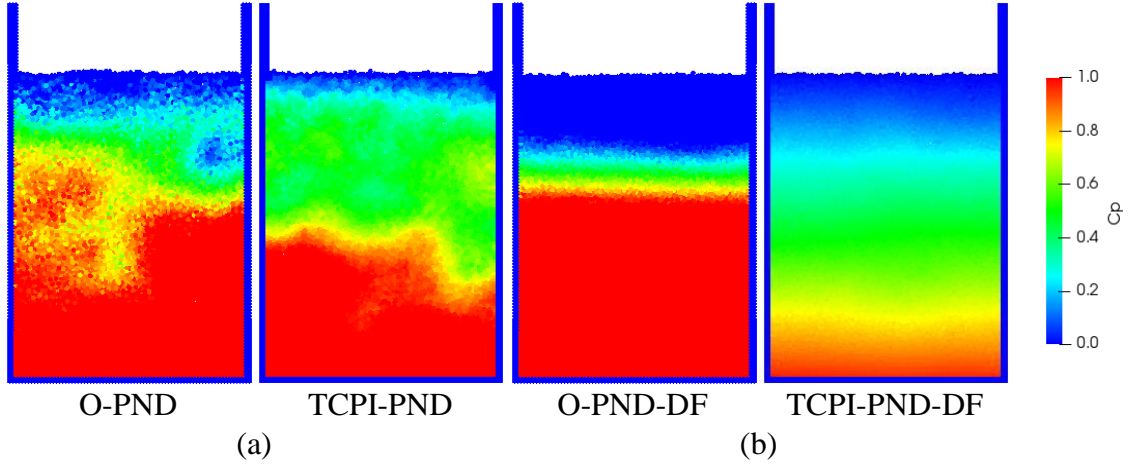

Figure 11: Hydrostatic pressure field at the instant 10 s. Source terms: (a) O-PND and TCPI-PND; (b) O-PND-DF and TCPI-PND-DF. Distance between particles $l_0 = 0.005\ m$ and time step $\Delta t = 1 \times 10^{-4}\ s$ (ratio $l_0/\Delta t = 50$).

In order to quantify the computed pressure oscillation, the standard deviations of the pressure coefficient ($Std_p$), see Eq. (46), for O-PND, O-PND-DF, TCPI-PND and TCPI-PND-DF in relation to the theoretical value ($C_P = 1$) are plotted in Figure 12. Figure 12(a) shows how the original source terms O-PND and O-PND-DF are highly influenced by the time step, and the oscillation increases remarkably as $\Delta t$ decreases. On the other hand, Figure 12(b) illustrates the very lower dependence of the proposed source terms TCPI-PND and TCPI-PND-DF in relation to the time step and the particle distance, showing remarkably improved stability. Also, much lower standard deviations were obtained for small time steps or larger particle distance, showing notable improvement achieved by the proposed source terms in extending the range of stable computation to higher time domain resolution or lower spatial resolution. The last one means more efficient numerical modeling with lower processing demand for similar stable results. Finally, in this pure hydrostatic case, the TCPI-PND-DF source term provided excellent results by mitigating the spurious pressure oscillations and improving the stability of the results with a combination of the time scale correction, zero variation of the particle number density and divergence-free conditions.

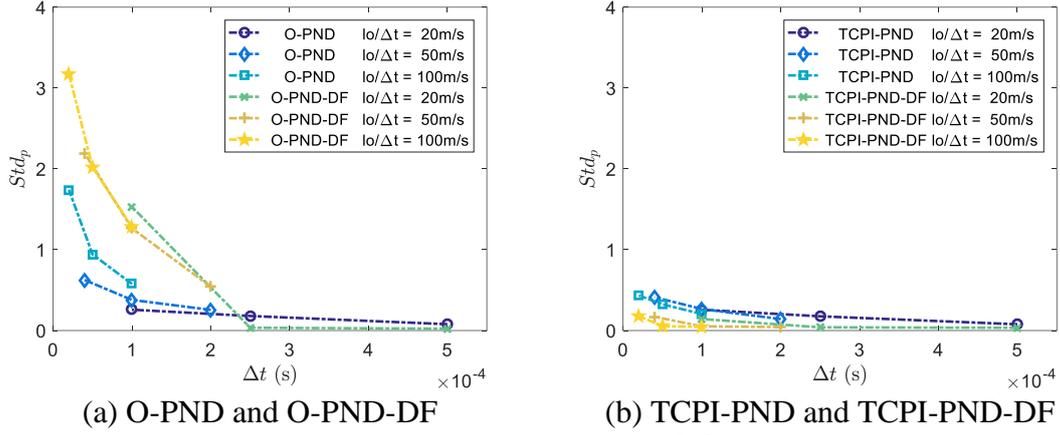

(a) O-PND and O-PND-DF

(b) TCPI-PND and TCPI-PND-DF

Figure 12: 2D hydrostatic tank. Standard deviation of pressure $Std_p$. (a) Original O-PND and O-PND-DF, (b) proposed TCPI-PND and TCPI-PND-DF source terms.

### 6.2. 2D water jet

Here a 2D case of hydrodynamic load due to a steady high velocity jet is considered. A flat wall is hit by perpendicular water jet of which the section width is $W_{jet} = 0.3$m, located $H_{jet} = 1.2$m above the wall of length $L_{wall} = 1.2$m. The inflow velocity is set to $V_{in} = -4$m/s. A schematic drawing of the main dimensions, pressure field and shape of the water jet after reach the steady state is shown in Figure 13. The fluid properties adopted herein are density of $\rho = 1000$kg/m$^3$ and kinematic viscosity of $\nu_k = 0.0$ (inviscid). The gravity effects are neglected so that the load is pure hydrodynamic one. Table 4 summarizes the numerical parameters used herein.

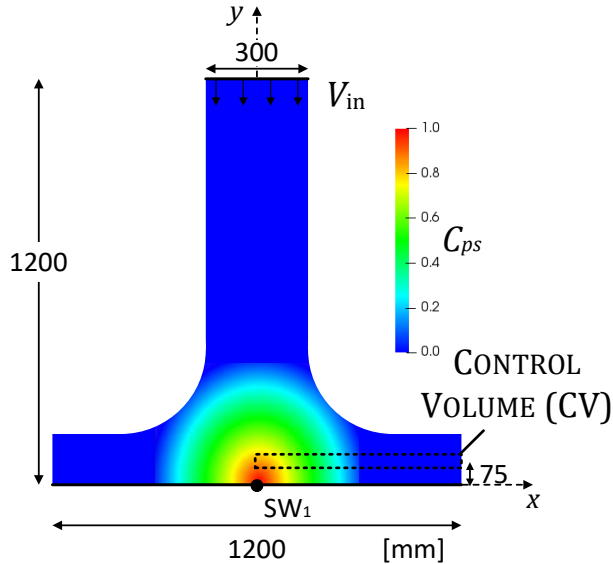

Figure 13: 2D water jet. Main dimensions, shape of the water jet after reach the steady state, sensor SW$_1$ at the position $(x, y) = (0,0)$ and control volume (CV). Pressure field in steady state flow.

The evolution of the non-dimensional pressure coefficient $C_{PS} = P/P_s$ computed at sensor SW$_1$ (see Figure 13) and the theoretical pressure coefficient $C_{PS} = 1$, represented by a blue solid line, are compared in the following analysis. $P_s = 0.5\rho V_{in}^2$ is the steady state analytical solution of the pressure at the stagnation point $(x, y) = (0,0)$, i.e., null velocity (Taylor, 1966). The dimensionless time $(\tau_u)$ is defined as $\tau_u = t|V_{in}|/H_{jet}$.

Figure 14 provides the time series of raw pressure coefficients computed by using O-PND and TCPI-PND source terms. The figure shows that computed pressure coefficients agree well with the analytical one. Similar to the previous results from the hydrostatic simulations, again higher pressure oscillations associated to the decrease of time step occur when the original formulation O-PND is applied, whereas a better performance is achieved by the application of the proposed source term TCPI-PND ensuring a lower pressure oscillation, almost independent of the time step.

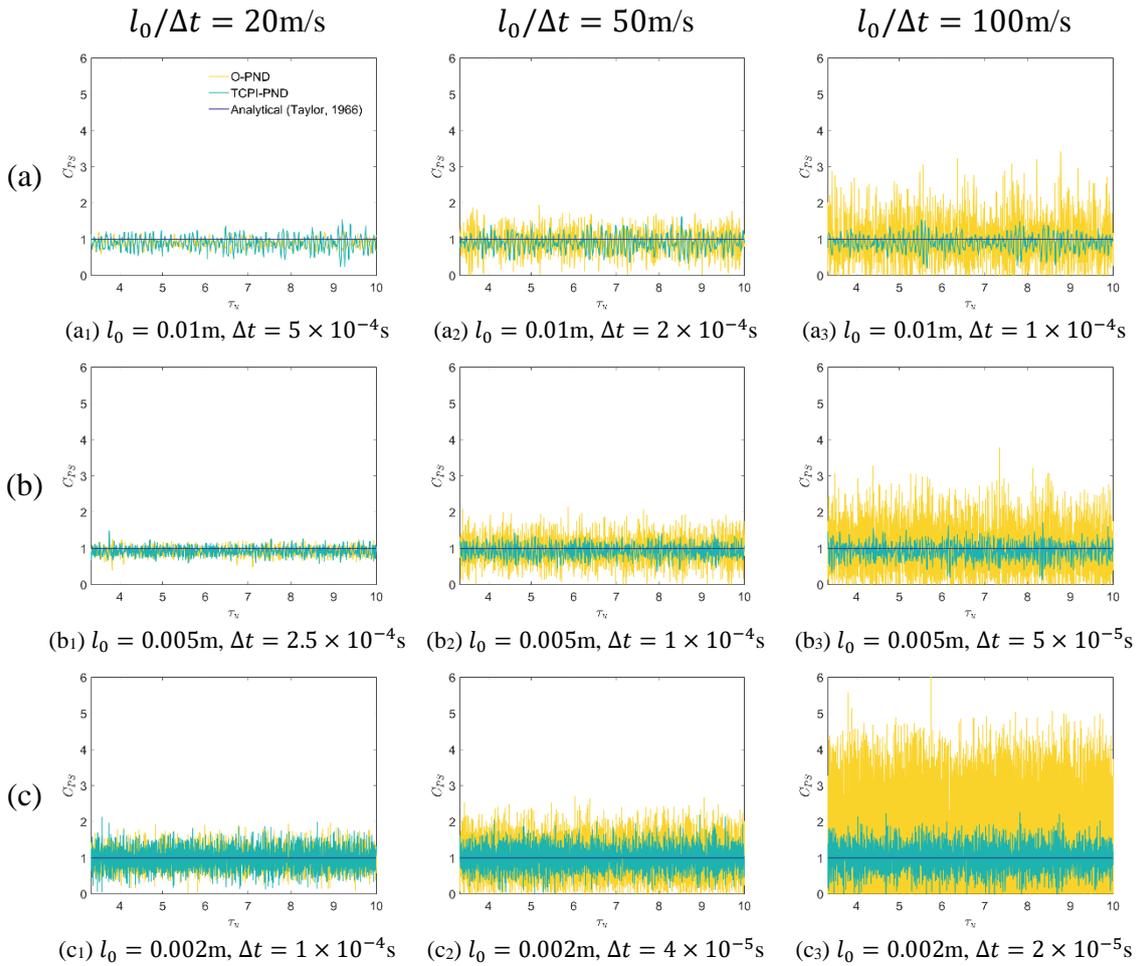

Figure 14: 2D water jet. Pressure at sensor SW$_1$. Source terms with the zero variation of the particle number density O-PND and TCPI-PND.

The time series of raw pressure coefficients computed by using O-PND-DF and TCPI-PND-DF source terms are presented in Figure 15. The amplitude of the spurious pressure oscillations computed by adopting O-PND-DF become remarkably higher for smaller time steps, while they are much more stable and almost independent of the time step when computed by using the TCPI-PND-DF source term. Moreover, the adoption of O-PND-DF do not only give higher spurious pressure oscillations but also cause unstable failed simulations for $l_0 = 0.005$ m with $\Delta t = 2.5 \times 10^{-4}$ s, and $l_0 = 0.002$ m with $\Delta t = 1 \times 10^{-4}$ s (see Figure 15(b$_1$) and (c$_1$)).

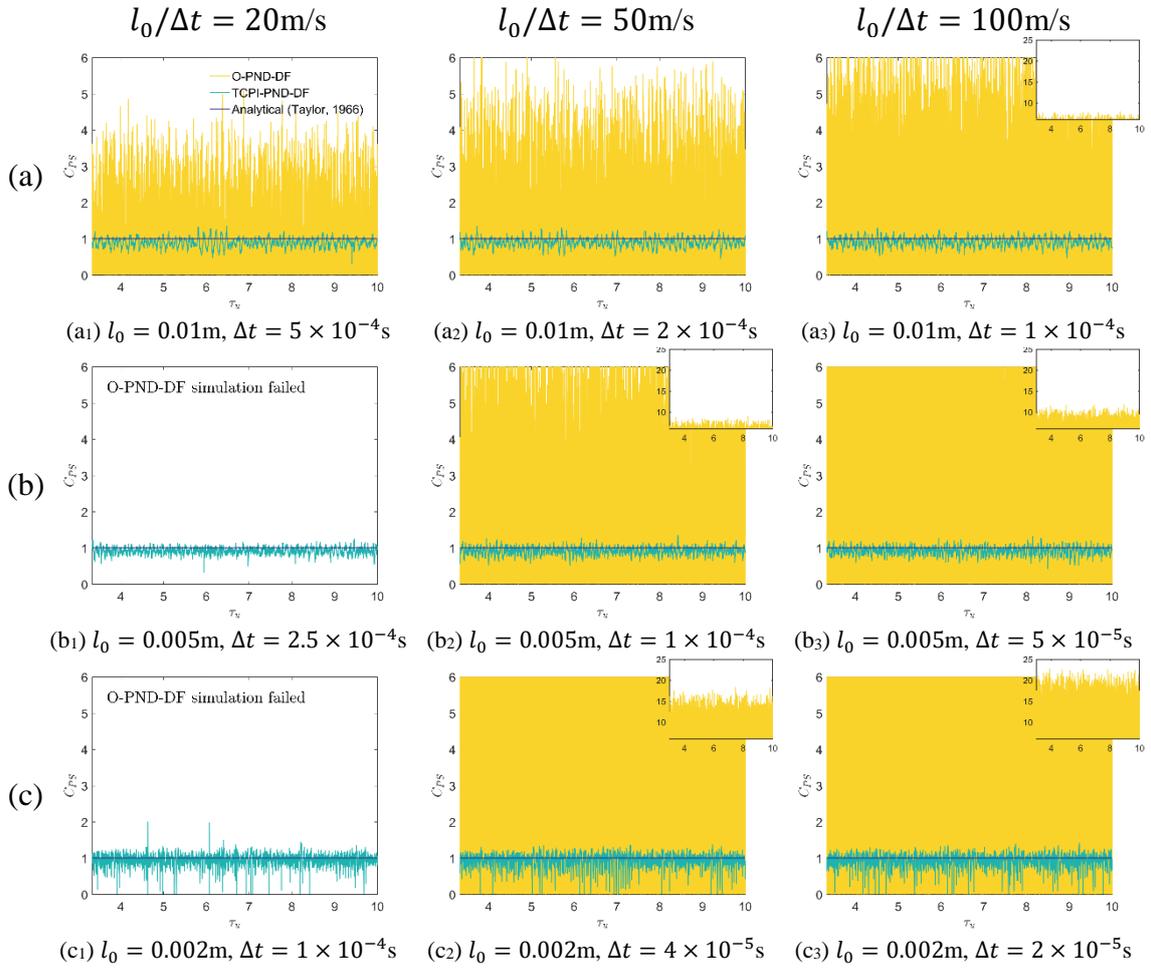

Figure 15: 2D water jet. Pressure at sensor SW$_1$. Source terms with the zero variation of the particle number density and velocity-divergence-free condition O-PND-DF and TCPI-PND-DF.

Figure 16 illustrates the computed pressure fields at the instants $t = 1.0$ and $2.0$ s ($\tau_u = 3.33$ and $6.66$). Also, the contour shape $(x_s, y_s)$ of a perpendicular water jet imping on a rigid wall obtained by the steady state analytical solution (Milne-Thomson, 1968) is marked by the orange squares:

$$y_s = L\left\{\frac{1}{2} + \frac{1}{\pi}\log\coth\left[\frac{\pi}{4}\left(\frac{2x_s}{L} - 1\right)\right]\right\}. \tag{51}$$

In Figure 16, half of the symmetric water jet geometry represents the result obtained by each of the source terms using $l_0 = 0.005$ m and $\Delta t = 5 \times 10^{-5}$ s (ratio $l_0/\Delta t = 100$ m/s). The simulations with original sources O-PND and O-PND-DF are characterized by unphysical pressure distributions and significant oscillations in time. The figure shows that all source terms capture the free surface solution in a satisfactory way. The adoption of the proposed approach results in improved pressure fields and mitigated the pressure oscillations in time. TCPI-PND source term leads to better pressure distribution, although there still exist some numerical pressure oscillation. The proposed TCPI-PND-DF source term provides significantly enhanced pressure field characterized by almost semi-ellipsoidal and distinctive smooth pressure contours at the high-pressure region in the vicinity of the stagnation point.

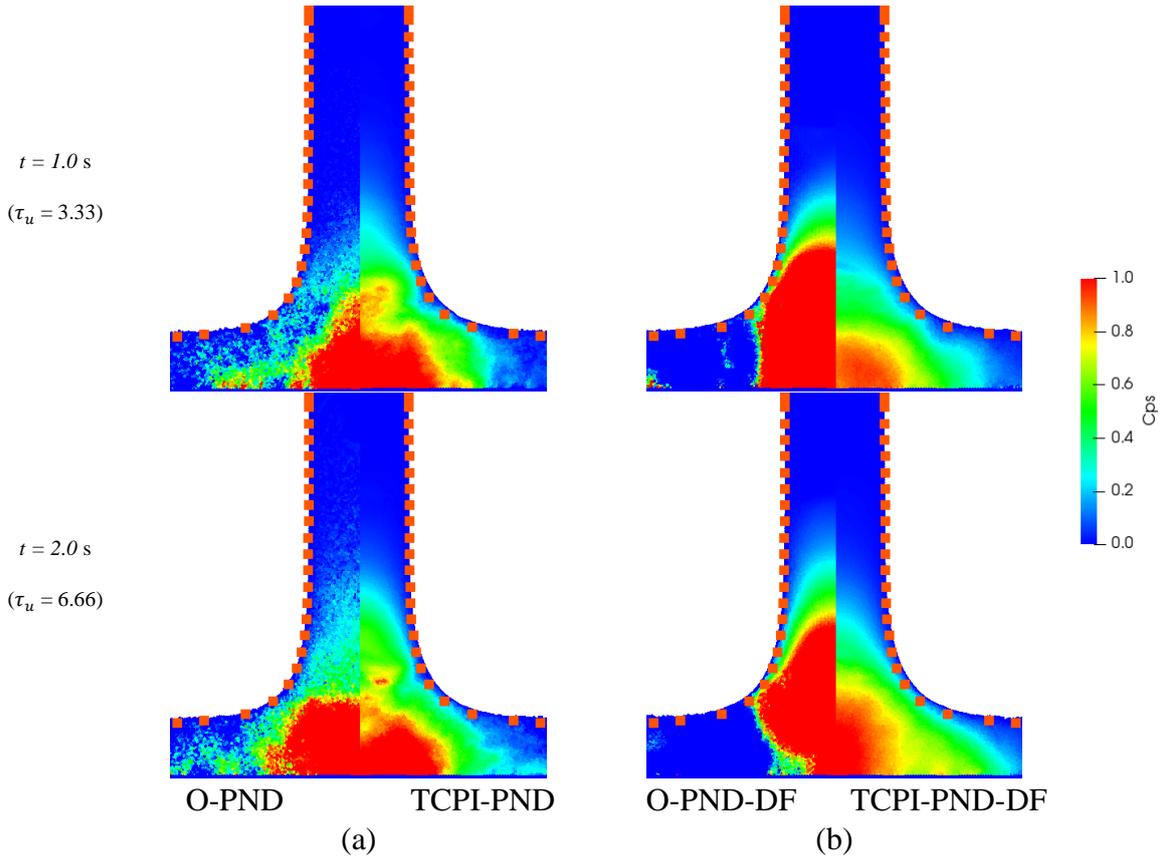

Figure 16: 2D water jet pressure field. Source terms: (a) O-PND and TCPI-PND; (b) O-PND-DF and TCPI-PND-DF. Distance between particles $l_0 = 0.005$m and time step of $\Delta t = 5 \times 10^{-5}$s (ratio $l_0/\Delta t = 100$). Evolution of water jet profile at times 1.0 and 2.0s ($\tau_u = 3.33$ and 6.66). The orange squares represent the free-surface steady state analytical solution (see Eq. (51)).

In order to show how the pressure is affected by time step, the standard deviation of pressure coefficient ($Std_{ps}$) between the instants $t = 1$s ($\tau_u = 3.33$) and 3s ($\tau_u = 10$), i.e., flow in steady state, is given in Figure 17. From the Figure 17(a), the standard deviations obtained with O-PND and O-PND-DF increase with reduction of the time step, while in Figure 17(b), the numerical results demonstrated that the pressure oscillations become almost independent to both time step and particle distance after the adoption of the proposed source terms TCPI-PND and TCPI-PND-DF. As already commented, the simulations for $l_0 = 0.005$ m with $\Delta t = 2.5 \times 10^{-4}$ s, and $l_0 = 0.002$ m with $\Delta t = 1 \times 10^{-4}$ s by using O-PND-DF were unstable and failed. Therefore, the standard deviations of these simulations are not computed and not plotted here.

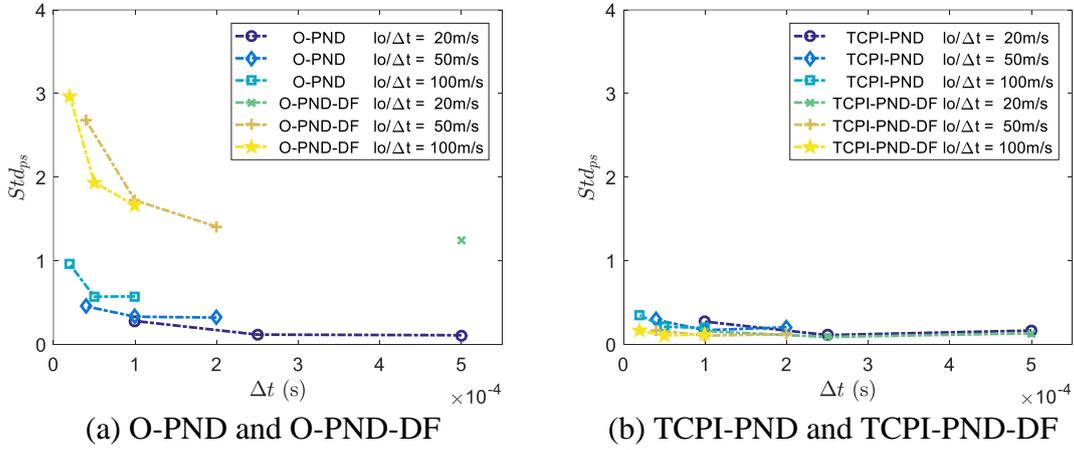

(a) O-PND and O-PND-DF  (b) TCPI-PND and TCPI-PND-DF
Figure 17: 2D water jet. Standard deviation of pressure ($Std_{ps}$). (a) Original O-PND and O-PND-DF, (b) proposed TCPI-PND and TCPI-PND-DF source terms.

Figure 18 gives an overview of the performances of the source terms in the pure hydrodynamic case. The diameter of the circles represents the magnitude of the standard deviations and cross markers denote unstable simulations. The results obtained by the original source terms are shown in blue and those obtained by the proposed sources terms are shown in orange color.

The direct comparison between the results of O-PND with TCPI-PND provided by Figure 18(a) shows the improvements achieved by the proposed approach, which leads to much more stable pressure computation as the discrete time step decreases. On the other hand, the direct comparison between the results of O-PND-DF with TCPI-PND-DF in Figure 18(b) shows clearly that, in this pure hydrodynamic case, the source term O-PND-DF produces the worst results, while TCPI-PND-DF achieved the most stable computations.

The grey background shaded in Figure 18(b) are used to identify the region the O-PND-DF source term fails. As already commented, the simulations for $l_0 = 0.005$ m with $\Delta t = 2.5 \times 10^{-4}$ s, and $l_0 = 0.002$ m with $\Delta t = 1 \times 10^{-4}$ s by using O-PND-DF were unstable and are indicated by the cross markers.

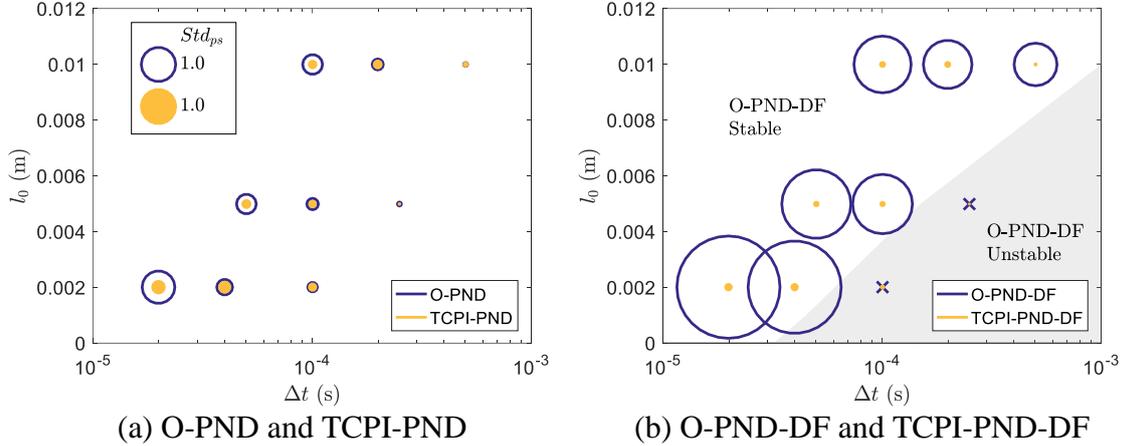

(a) O-PND and TCPI-PND   (b) O-PND-DF and TCPI-PND-DF

Figure 18: 2D water jet. Standard deviation of pressure ($Std_{ps}$) in spatial-time region. The magnitude of Std is represented by the circle's diameter and cross markers represent failed simulations. (a) O-PND and TCPI-PND, (b) O-PND-DF and TCPI-PND-DF source terms. Underlying colors are used to identify the region in which the O-PND-DF computations failed.

To check the convergence of the different source terms, the mean value of the computed stagnation pressure between the instants $t = 1$ s ($\tau_u = 3.33$) and 3 s ($\tau_u = 10$), i.e., flow in steady state, are provided as a function of the particle distance $l_0$ in Figure 19. As a reference, the analytical solution of the pressure, non-dimensional pressure $C_{ps} = 1$ is also shown. According to Figure 19(a), the computed values using original O-PND and O-PND-DF source terms converge to the analytical solution as $l_0$ goes to zero, except for O-PND simulations with ratio $l_0/\Delta t = 100$ m/s, and unstable and failed O-PND-DF simulations using $l_0 = 0.005$ m with $\Delta t = 2.5 \times 10^{-4}$ s, and $l_0 = 0.002$ m with $\Delta t = 1 \times 10^{-4}$ s. The convergence of the proposed TCPI-PND and TCPI-PND-DF source terms is presented in Figure 19(b), which shows that the computed value tends monotonically to the analytical one for all simulations.

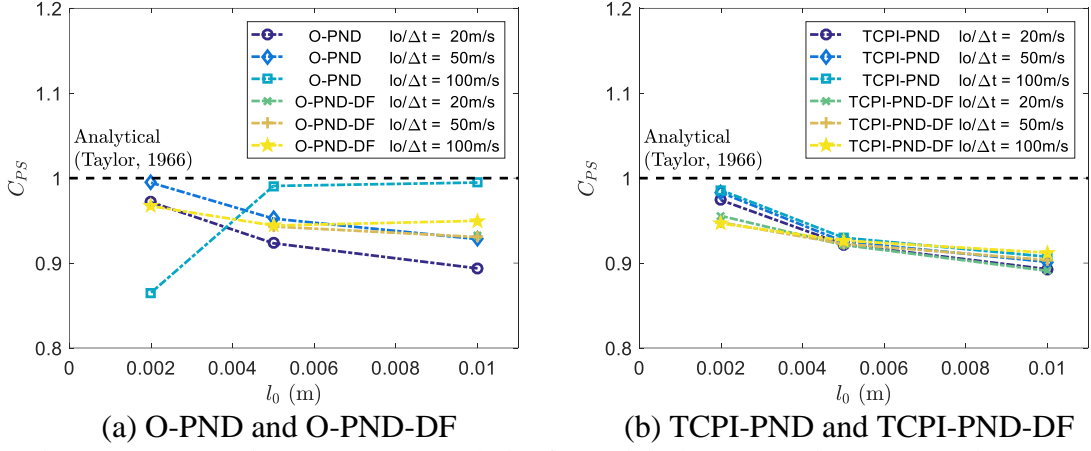

(a) O-PND and O-PND-DF        (b) TCPI-PND and TCPI-PND-DF

Figure 19: 2D water jet. Convergence analysis of (a) Original O-PND and O-PND-DF, (b) proposed TCPI-PND and TCPI-PND-DF source terms.

Finally, in order to verify the influence that the improved source terms have on the velocity field, the horizontal velocities $u$ computed at the instant $t = 2.0$s (steady-state) for all particles within the rectangular control volume (CV) of length $L_{jet}/2 = 0.6$m and height $2l_0$ (see Figure 13) were compared against the analytical solution of velocity on the wall $u_{wall}$. In case of jet impinging on a perpendicular flat plate, the velocity was obtained analytically by solving the implicit system (Taylor, 1966):

$$\frac{u_{wall}}{V_{in}} = \frac{\sqrt{1-q^2}-1}{q} \quad , \quad 0 < q < 1 \tag{52}$$

and

$$\frac{x_{wall}}{W_{jet}} = \frac{1}{2\pi}\log\left(\frac{1+q}{1-q}\right) + \frac{1}{\pi}\sin^{-1}q \quad , \quad 0 < q < 1 \tag{53}$$

where $V_{in}$ stands for the inflow velocity, $x_{wall}$ is a coordinate on the flat plate, $W_{jet}$ denotes the width of the water jet and $q$ is an auxiliary variable.

Since the no-slip condition was adopted for the boundary walls in the present MPS, the computed velocity is zero at the wall and increases with the perpendicular distance from the wall. Therefore, instead the computed velocities on the wall, the velocities computed for fluid particles within the fixed control volume 0.075m away from the wall (see Figure 13) were compared with the idealized analytical solution of velocity on the wall given by Eq. (52), as depicted in Figure 20. The results using the original source terms O-PND and O-PND-DF capture the velocity distribution solution in a satisfactory way, but the simulations using O-PND show notable numerical instabilities and convergence is not achieved, as shown in Figure 20(a). The computed velocities using the proposed source terms TCPI-PND and TCPI-PND-DF converge to the analytical solution as the particle distance $l_0$ decreases and are more accurate and smoother than the ones obtained using the original source term O-PND, as demonstrated by smaller oscillations closer to the analytical solution. Notwithstanding, the velocities are slight underestimated by the computed ones, although a higher resolution should provide better results, since they converge towards the analytic solution as the resolution increases, or even for coarse resolutions the adoption of higher order differential operators (Randles & Libersky, 1996; Suzuki, 2008; Tamai & Koshizuka, 2014; Duan et al., 2018; Liu et al., 2018; Garoosi & Shakibaeinia, 2021) enhances the energy conservation (see Appendix A) and consequently reproduces better the velocity field.

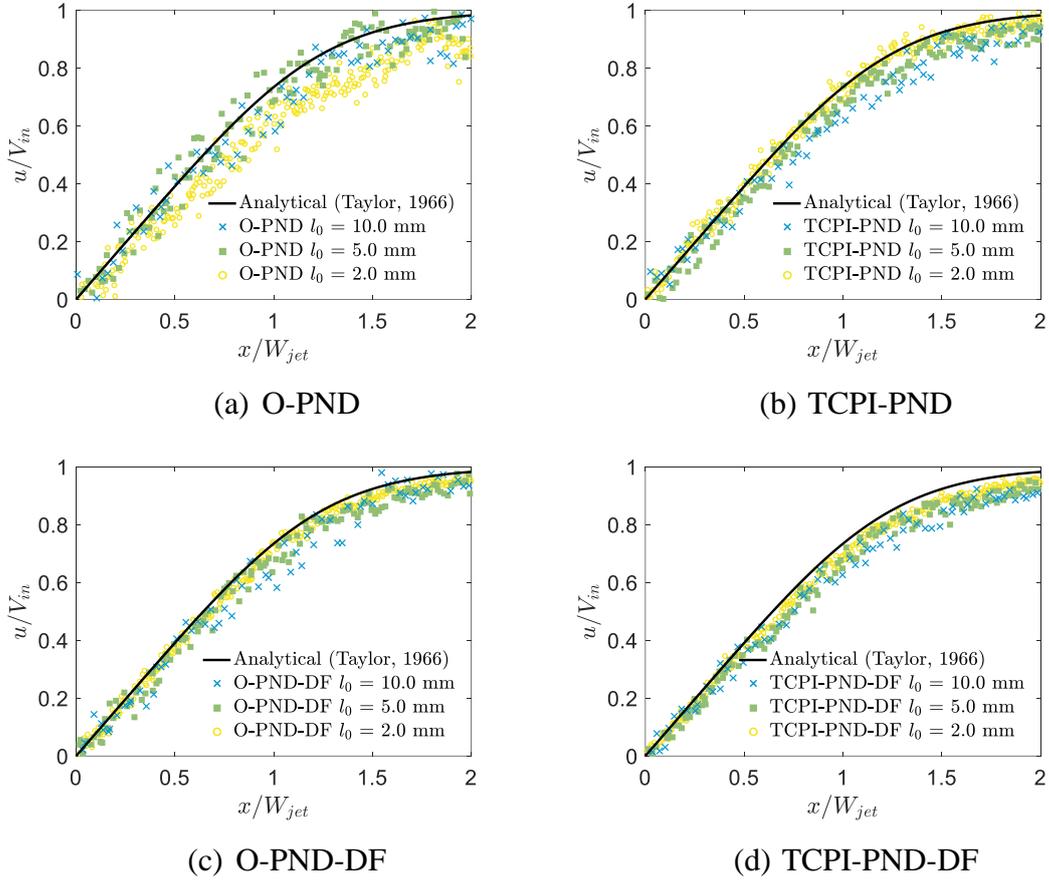

Figure 20: 2D water jet. Horizontal velocity $u$ normalized by the jet inflow velocity $V_{in}$ along the horizontal distance $x$ normalized by the width of the water jet $W_{jet}$. (a) Original O-PND and (b) O-PND-DF, and proposed (c) TCPI-PND and (d) TCPI-PND-DF source terms. Simulations using the particle distances $l_0 = 0.01, 0.005,$ and $0.002$m, and time step $\Delta t = l_0/100$.

6.3. **2D dam breaking**

In order to verify the performance of the proposed source terms concerning the computation of transient hydrodynamic impact loads, the numerical simulation of a 2D dam-break flow is carried out. Figure 21 shows the main dimensions of the tank of height $H_T = 0.6$m and length $L_T = 1.61$m, and the water column of height $H_F = 0.3$m, length $L_F = 0.6$m (Lobovský et al., 2014) and initial hydrostatic pressure[1] (Pohle, 1951; Cao et al., 2018).

The fluid properties are density of $\rho = 997$kg/m$^3$ and kinematic viscosity of $\nu_k = 0.89 \, x \, 10^{-6}$m$^2$/s. The pressures computed by the original and proposed source terms are compared against experimentally measured at sensor SD$_1$ at bottom corner of the opposite side wall, 3mm height from the bottom. The numerical parameters presented in Table 4 are adopted for all simulations.

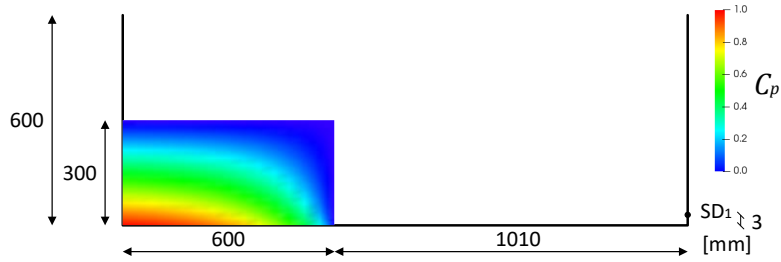

Figure 21: 2D dam breaking. Main dimensions and sensor position (Lobovský et al., 2014). Pressure field in the initial configuration[1].

The time histories of non-dimensional pressure coefficient $C_P = P/\rho g H$ at sensor SD$_1$, without any filtering, are shown in Figure 22. The dimensionless time ($\tau$) is defined as $\tau = t\sqrt{g/H_F}$.

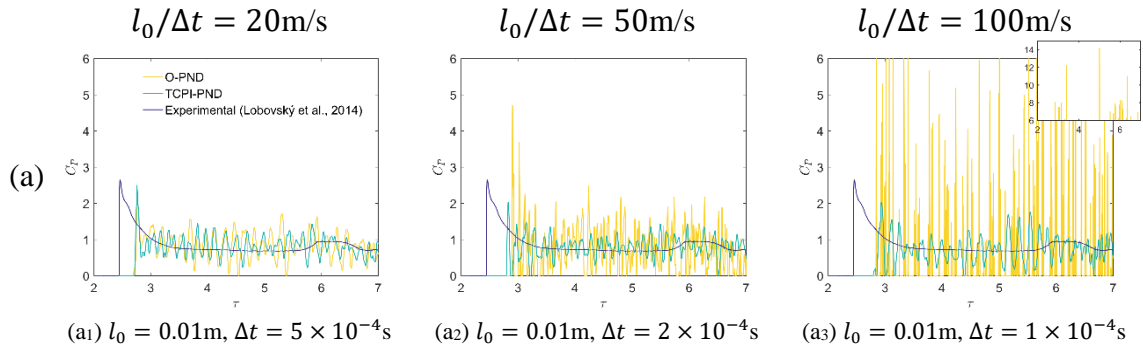

(a)

(a$_1$) $l_0 = 0.01$m, $\Delta t = 5 \times 10^{-4}$s    (a$_2$) $l_0 = 0.01$m, $\Delta t = 2 \times 10^{-4}$s    (a$_3$) $l_0 = 0.01$m, $\Delta t = 1 \times 10^{-4}$s

---

[1] The hydrostatic pressure can be computed as $P_0(x,y) = \rho g(H - y) - \frac{8\rho g H}{\pi^2}\sum_{m=0}^{100}\left[\frac{1}{(2m+1)^2}e^{\frac{(2m+1)\pi x}{2H}}cos\left(\frac{2m+1}{2H}\pi y\right)\right]$.

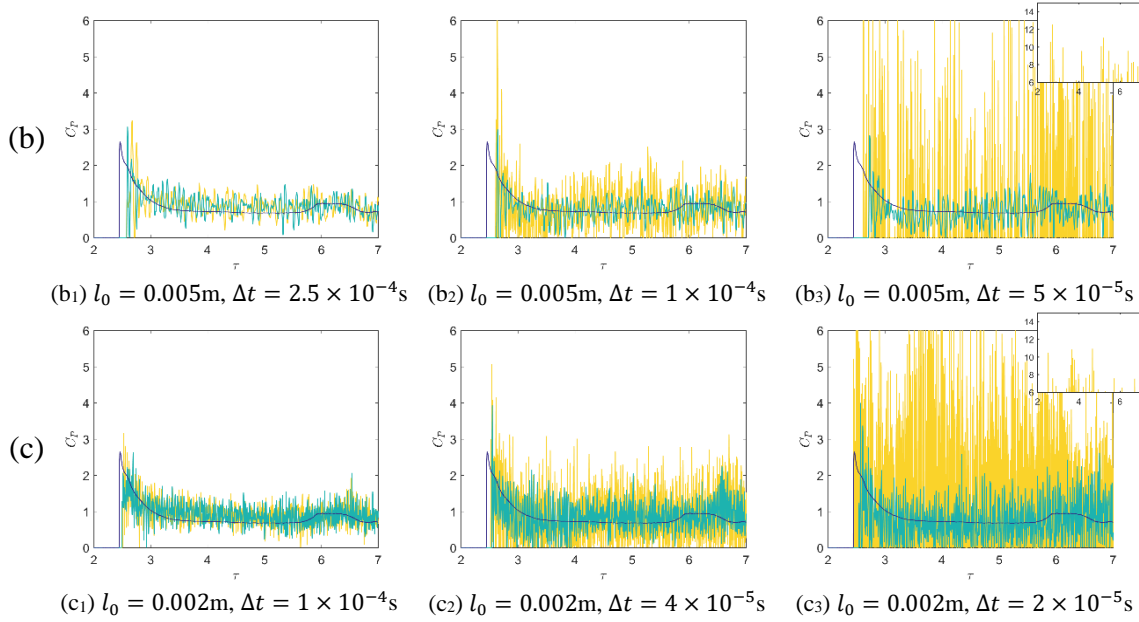

(b₁) $l_0 = 0.005$m, $\Delta t = 2.5 \times 10^{-4}$s  (b₂) $l_0 = 0.005$m, $\Delta t = 1 \times 10^{-4}$s  (b₃) $l_0 = 0.005$m, $\Delta t = 5 \times 10^{-5}$s

(c₁) $l_0 = 0.002$m, $\Delta t = 1 \times 10^{-4}$s  (c₂) $l_0 = 0.002$m, $\Delta t = 4 \times 10^{-5}$s  (c₃) $l_0 = 0.002$m, $\Delta t = 2 \times 10^{-5}$s

Figure 22: 2D dam breaking. Pressure at sensor SD$_1$. Source terms with the zero variation of the particle number density O-PND and TCPI-PND.

According to Figure 22, the decrease of time step, i.e., increase of the ratio $l_0/\Delta t$, increases the amplitude of pressure oscillations computed by using O-PND. On the other hand, relatively lower oscillations, whose magnitudes are almost independent to the numerical parameters, were obtained by using TCPI-PND, following the same tendency obtained in the previous simulations for the hydrostatic (see Figure 9(a₃), (b₃) and (c₃)) and the pure hydrodynamic cases (see Figure 14 (a₃), (b₃) and (c₃)).

The dam-break problem is also simulated by using O-PND-DF and TCPI-PND-DF source terms. Figure 23 depicts the time histories of raw pressure coefficient at sensor SD$_1$ computed by source terms O-PND-DF and TCPI-PND-DF, which are presented together with the pressure measured in the experiment (Lobovský et al., 2014). Again, the decrease of time step, keeping the particle distance constant, produces higher pressure oscillations computed by O-PND-DF, while the pressure oscillations computed by TCPI-PND-DF are much smaller. Furthermore, within the relatively wide range of the numerical parameters analyzed herein, the magnitude of pressure oscillations computed by TCPI-PND-DF are essentially independent of time step and particle distance, showing more stable results. Moreover, since the propagation speed used in the former hydrostatic case was also adopted in this case, the results show that the straightforward calibration of $c_s$ also works relatively well for this dynamic case.

Moreover, Figure 22 and Figure 23 show that as the particle model resolution increases, the instant of the computed first peak pressure agree well with the experiment, i.e., the numerical solution converges to the experimental one, whereas a delay occurs for the coarse resolutions.

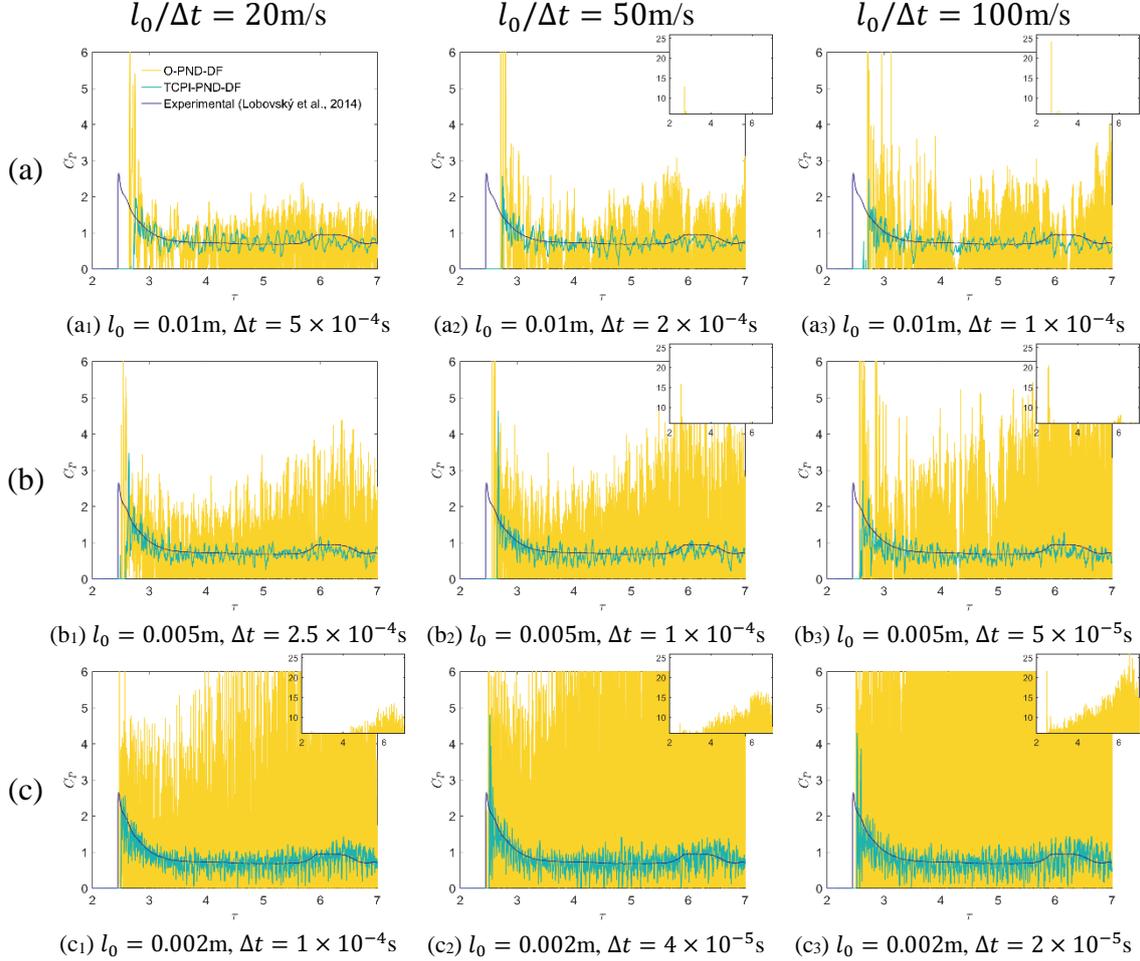

Figure 23: 2D dam breaking. Pressure at sensor $SD_1$. Source terms with the zero variation of the particle number density and velocity-divergence-free condition O-PND-DF and TCPI-PND-DF.

Figure 24 illustrates the snapshots showing free-surface profile of the collapsing water column obtained from the simulations using the O-PND and TCPI-PND, whereas Figure 25 presents the snapshots from O-PND-DF and TCPI-PND-DF source terms, all with distance between particles of $l_0 = 0.005$ m and time step of $\Delta t = 5 \times 10^{-5}$ s (ratio $l_0/\Delta t = 100$ m/s). The color scale is associated to pressure magnitude. The pressure field near the stagnation point is also highlighted. The simulation carried out using O-PND leads to a rough pressure field with numerical oscillations and disordered particles. Although some unphysical pressure oscillations remained in the TCPI-PND simulations, smoother pressure field were obtained (see Figure 24 (b)). The significant improvement achieved by adopting TCPI-PND-DF (see Figure 25(b)) is evidenced by continuous and smooth pressure fields over the unphysically oscillating pressure fields obtained using O-PND or O-PND-DF.

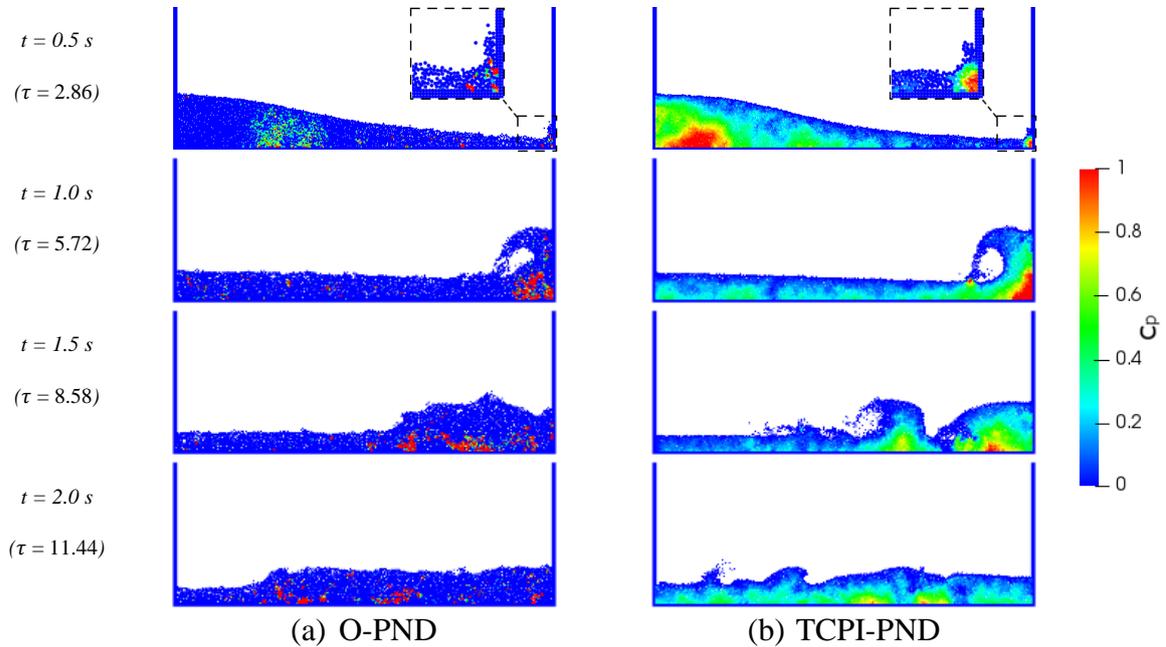

Figure 24: 2D dam breaking pressure field. (a) Original O-PND and (b) proposed TCPI-PND source terms. Distance between particles of $l_0 = 0.005$m and time step of $\Delta t = 5 \times 10^{-5}$s. Evolution of free-surface profile at times 0.5, 1.0, 1.5 and 2.0s ($\tau$ = 2.86, 5.72, 8.58 and 11.44).

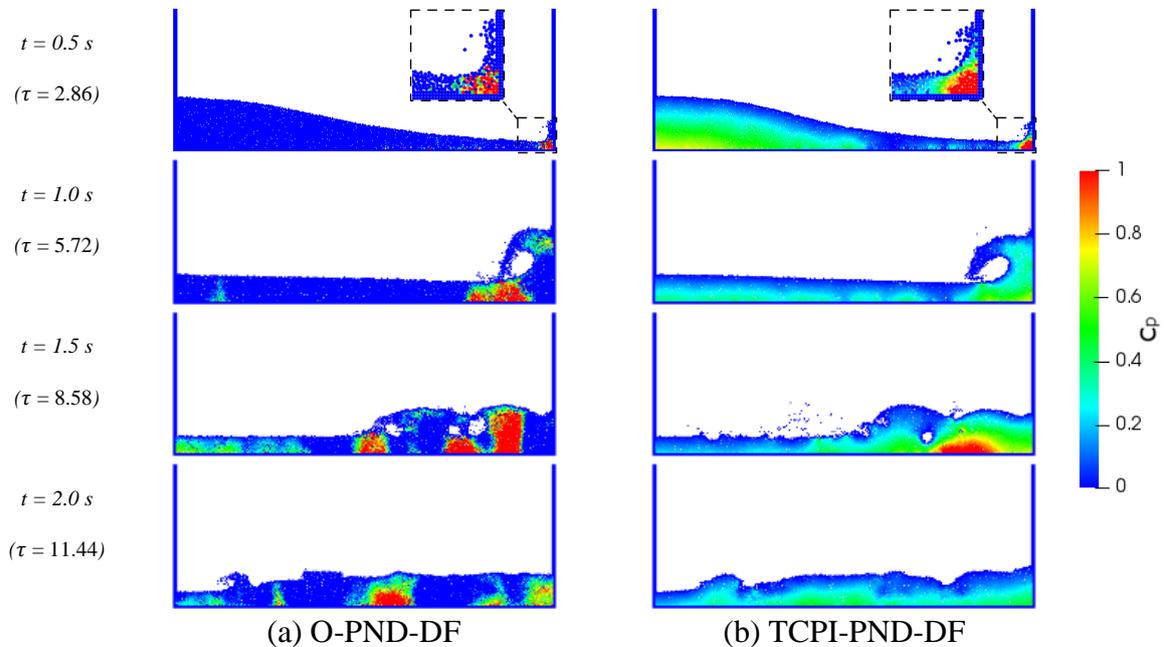

Figure 25: 2D dam breaking pressure field. (a) Original O-PND-DF and (b) proposed TCPI-PND-DF source terms. Distance between particles of $l_0 = 0.005$m and time step of $\Delta t = 5 \times 10^{-5}$s. Evolution of free-surface profile at times 0.5, 1.0, 1.5 and 2.0s ($\tau$ = 2.86, 5.72, 8.58 and 11.44).

In order to quantify the pressure oscillation and the error in relation to the experimentally measured pressure, the normalized root mean square deviation ($NRMSD_p$) was calculated between the instants $\tau_1 = 3.0$ and $\tau_2 = 5.0$ and the results are shown in Figure 26. The $NRMSD_p$ is defined here as:

$$NRMSD_p = \frac{RMSD_p}{|\bar{p}_n|} = \frac{1}{|\bar{p}_n|}\sqrt{\frac{1}{m}\sum_{i=1}^{m}|p_{n,i} - p_{e,i}|^2}, \tag{54}$$

where $RMSD_p$ denotes the root mean square deviation, $\bar{p}_n$ is the mean of the computed numerical results $p_n$, $p_e$ stands for the experimental results and $m$ is the number of values computed during a time interval.

Figure 26(a) shows that in the case of dam breaking, the magnitude of computed pressure oscillations ($NRMSD_p$) obtained by O-PND is almost constant for a fixed $l_0/\Delta t$, when $l_0/\Delta t \leq 50$m/s. According to the figure, increasing $l_0/\Delta t$ by decreasing time step or keeping time step and increasing the particle distance leads to higher computed pressure oscillations. On the other hand, Figure 26(b) shows that the influence of $l_0/\Delta t$, as well as time step and particle distance on pressure oscillations computed by using TCPI-PND is almost negligible. For all the cases analyzed herein, which cover a wide range of the numerical modeling parameters, the magnitude of pressure oscillations ($NRMSD_p$) computed by TCPI-PND is as small as the best result achieved by O-PND.

Regarding the source terms with the zero variation of the particle number density and velocity-divergence-free condition (O-PND-DF and TCPI-PND-DF) the pressure oscillations computed by O-PND-DF are very strongly influenced by the time step, meanwhile the results obtained by TCPI-PND-DF are much more stable and almost independent to both time step and particle distance.

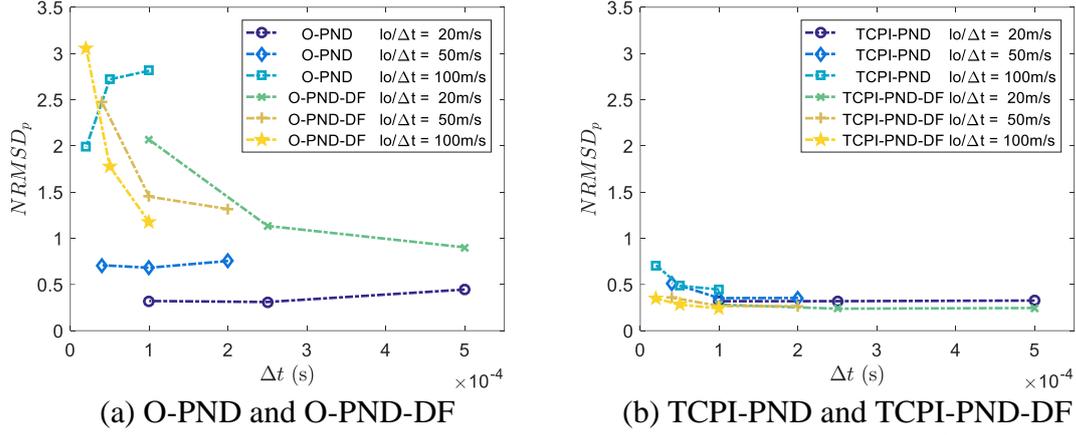

(a) O-PND and O-PND-DF  (b) TCPI-PND and TCPI-PND-DF

Figure 26: 2D dam breaking. Normalized root mean square deviation of pressure ($NRMSD_p$). (a) Original O-PND and O-PND-DF, (b) proposed TCPI-PND and TCPI-PND-DF source terms.

### 6.4. 3D sloshing in prismatic tank

The last case consists of 3D sloshing in a prismatic tank under translational coupled surge-sway (X and Y axis) motions. Figure 27 gives the geometry and the dimensions of the prismatic tank with 45 degrees chamfers in the top and bottom edges. The main internal dimensions of the tank are height $H_T = 0.54$m, width $W_T = 0.84$m and length $L_T = 0.72$m. In order to avoid hydroelastic effects, the tank was manufactured with Plexiglass of thickness $e_T = 30$mm. The performance of the proposed approach for the assessment of violent impact loads was evaluated for a filling ratio of 50% ($H_F = 0.27$m). The periods of surge and sway excitations are both selected to be $T_s = 1.25$s with amplitude motions of $A_x = 0.0144$m (0.02 x length) and $A_y = 0.0168$m (0.02 x width). The experiment was carried out by the authors, a hexapod-type motion platform (eMove eM6-400-1500) with six actuators was used to generate the motions, the dynamic pressure at P1 was measured by a piezoelectric pressure sensor (Kistler 211B6) and the sloshing wave motions were captured by a video camera.

In the numerical simulations, an initial particle spacing $l_0 = 0.01$m and four time steps $\Delta t = 5.0, 2.0, 1.0$ and $0.5 \times 10^{-4}$s were adopted. The fluid properties are density $\rho = 1000$kg/m$^3$ and kinematic viscosity of $\nu_k = 1.0 \; x \; 10^{-6}$m$^2$/s. The relaxation parameter $\gamma = 0.01$ and the propagation speed of the perturbations $c_s = 2$m/s were used herein. The simulations were carried out up to 20s, i.e., 16 cycles of excitations.

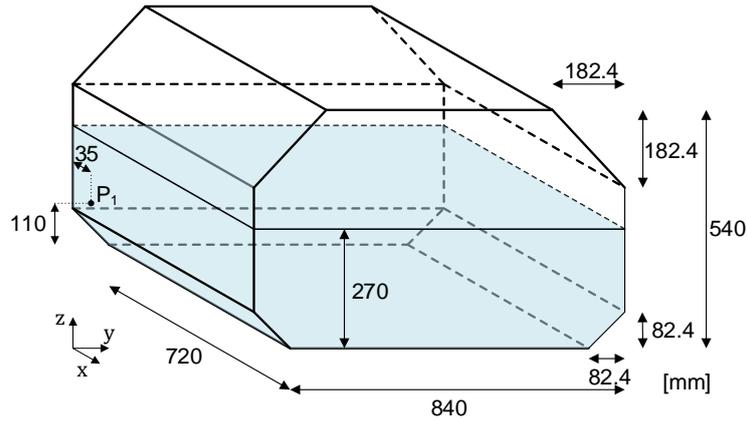

Figure 27: Schematic view of 3D sloshing. Main dimensions of tank, filling height and sensor position.

The non-dimensional pressure coefficient $C_P = P/\rho g H$ at P1 measured experimentally and computed numerically are shown in Figure 28. Both the experimental and numerical results consist of raw data, along the dimensionless time defined as $\tau_s = t/T_s$, obtained from the measurements and computations without applying any filtering technique. The results show that the general behaviors of the source terms regarding the time-stability of the pressure computation in the 3D cases are similar to the former 2D cases. For $\Delta t \geq 1 \times 10^{-4}$s, pressures computed by O-PND and TCPI-PND present similar oscillations. Nevertheless, when the time step is reduced to $5 \times 10^{-5}$ s, much higher pressure oscillations were computed by O-PND than TCPI-PND. On the other hand, significant improvements have been achieved by adopting TCPI-PND-DF, which lead to a very robust computation due to its more stable results regardless of the variation of the numerical parameters. This robustness is followed by TCPI-PND, and then original formulations O-PND and O-PND-DF, with the last one being the worst one because it is very sensitive to the reduction of time step. In addition to this, the pressures computed by TCPI-PND-DF, which include the impact loads, are in good agreement with the experimental result, even for the coarse particle distance adopted here.

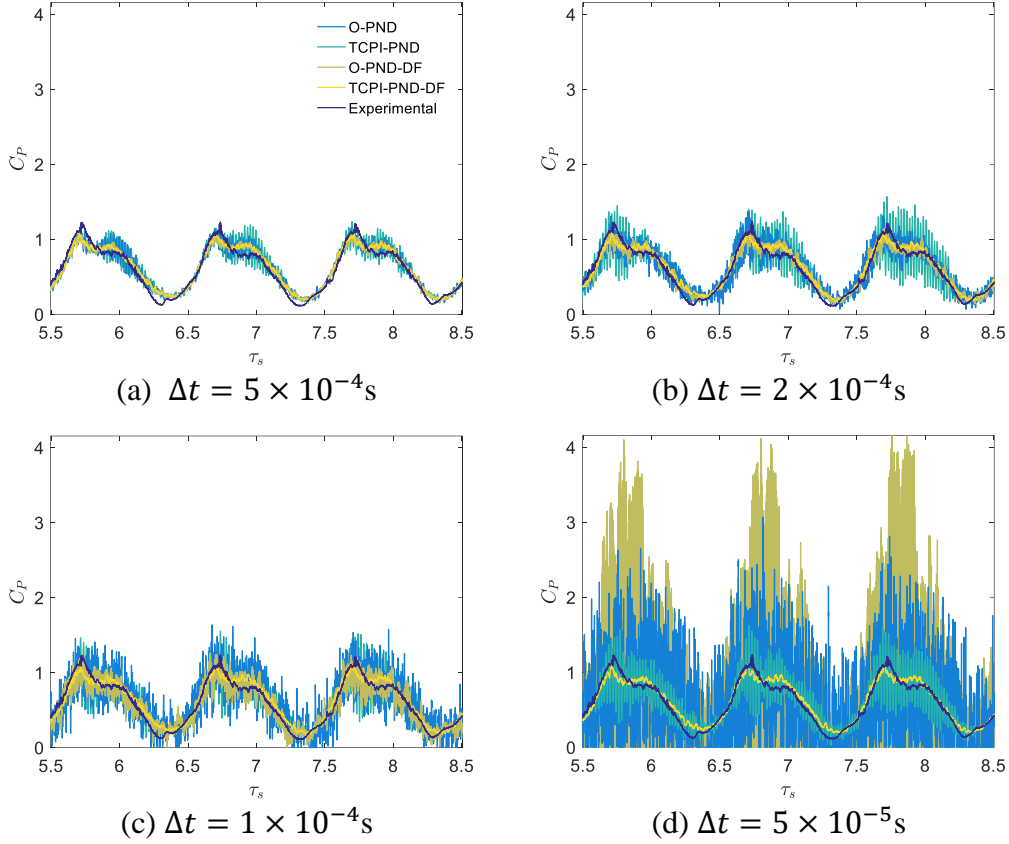

Figure 28: 3D sloshing in prismatic tank. Pressure at sensor $P_1$. Source terms with the zero variation of the particle number density (O-PND and TCPI-PND) and velocity-divergence-free condition (O-PND-DF and TCPI-PND-DF). Cases simulated with: $l_0 = 0.01\ m$ and (a) $\Delta t = 5 \times 10^{-4}$, (b) $\Delta t = 2 \times 10^{-4}$, (c) $\Delta t = 1 \times 10^{-4}$, and (d) $\Delta t = 5 \times 10^{-5}$s.

Sloshing wave profiles at different instants obtained experimentally and computed by using O-PND and TCPI-PND source terms with time step of $1 \times 10^{-4}$s are presented in Figure 29. It is important to point out that in the numerical results, the pressure distributions on the walls and velocity field on the free surface are also shown according to the color scales. Both simulations capture the highly deformed sloshing waves involving fluid fragmentation and merging. The superposition of the wave components in the length and width directions of the tank generates a swirling wave motion. The swirling wave hits one corner of the tank at the instants $\tau_s = 5.40, 7.20$ and $11.20$, whereas another corner is hit at $\tau_s = 6.80$ and $9.72$, approximately. At the instant $\tau_s = 8.48$, the splashing collapses and impacts on the main water body. Compared to O-PND, the TCPI-PND-DF simulation has substantially reduced pressure fluctuations, thereby providing much more smooth and realistic pressure distribution. A video with the 3D sloshing experiment and numerical simulations is also provided in the supplementary data (Video S1).

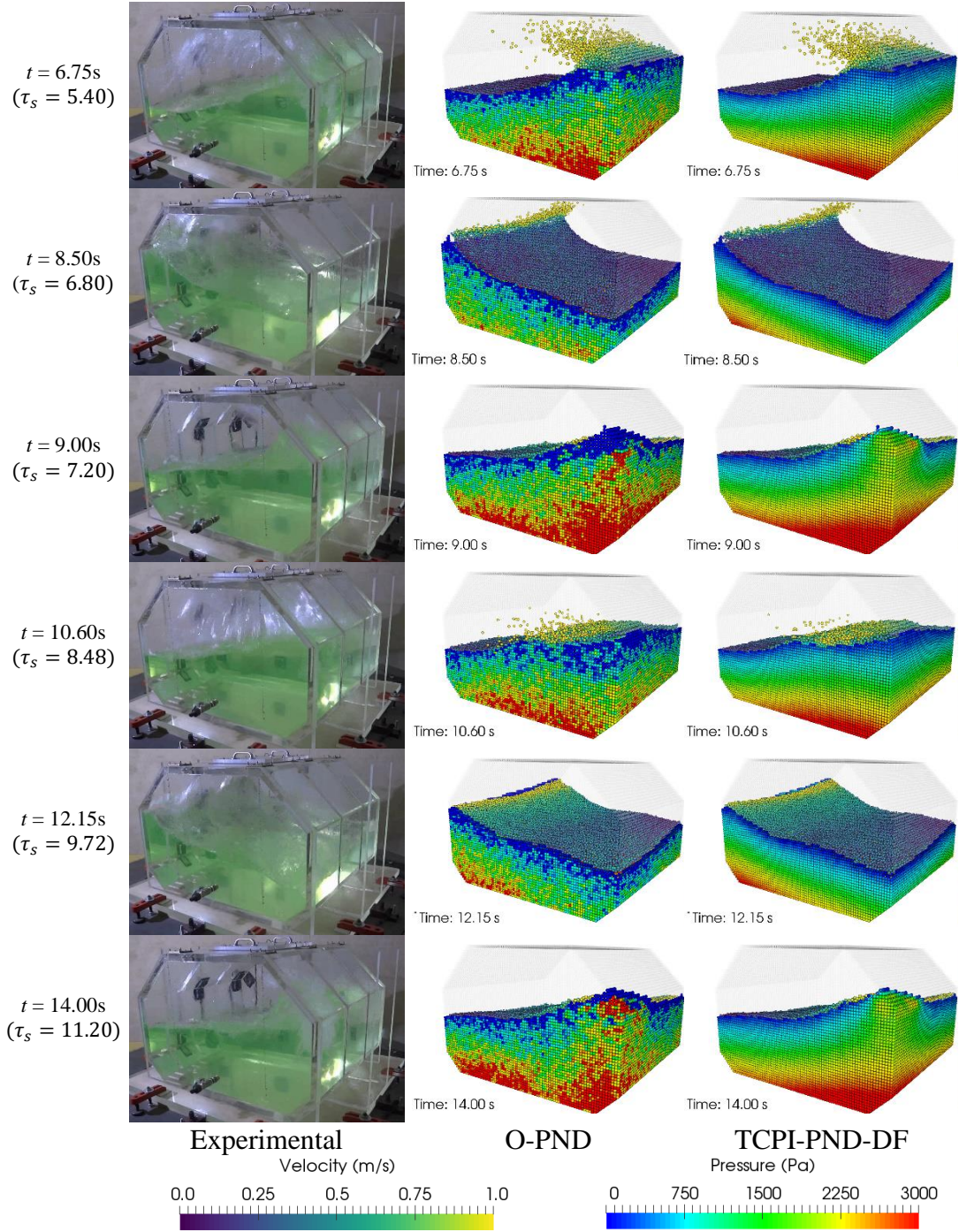

Figure 29: Experimental and numerical simulations of 3D sloshing. Original O-PND and proposed TCPI-PND-DF source terms. Distance between particles of $l_0 = 0.01\ m$ and time step of $\Delta t = 1 \times 10^{-4}\ s$ (ratio $l_0/\Delta t = 100$). Evolution of free-surface profile at times 6.75, 8.50, 9.00, 10.60, 12.15 and 14.0 s ($\tau_s$ = 5.40, 6.80, 7.20, 8.48, 9.72 and 11.20). The color scale of free surface particles indicates velocity magnitude, and the color scale of the wall particles indicates pressure. A video with the experiment and numerical simulations can be found in the supplementary material (Video S1).

Figure 30 illustrates the normalized root mean square deviations ($NRMSD_p$), see Eq. (54), calculated between the instants $\tau_s = 6.4$ and 7.4 (7th cycle). Both original source terms O-PND and O-PND-DF are influenced by the time step. Especially for $\Delta t \leq 1 \times 10^{-4}$ s, the computed pressures become very unstable with very large oscillation magnitudes. In contrast, as in the previous 2D simulations, the proposed source terms TCPI-PND and TCPI-PND-DF lead to a very lower dependence in relation to the time step. For O-PND with fine-tuned relaxation parameter $\gamma$, the magnitude of pressure oscillation might be slightly lower than TCPI-PND when relatively larger time steps are adopted. However, it is important to emphasize that with such large time steps generally the numerical stability criteria (Eq. (24)) might be a concern for the computation. The O-PND-DF source terms provides slightly better results than O-PND and TCPI-PND, but it becomes highly unstable when small time steps are used. Finally, it is interesting to point out the lower pressure oscillation obtained by TCPI-PND-DF compared to another source terms, indicating that for this specific situation, with high-impulsive cyclic loads, the adoption of the PND deviation and divergence-free condition is highly recommended.

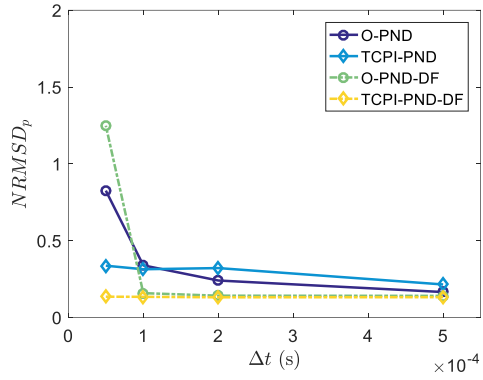

Figure 30: 3D sloshing in prismatic tank. Normalized root mean square deviation of pressure ($NRMSD_p$). Original O-PND and O-PND-DF, and proposed TCPI-PND and TCPI-PND-DF source terms. Distance between particles of $l_0 = 0.01$m.

## 7. Concluding Remarks

In the present study, the spurious numerical oscillations associated to the Lagrangian particle-based simulations are analyzed through the time-stability issue of the computed pressure, and an original interpretation based on the numerical modeling issues of the particle-level collision/impact is presented. By considering the momentum conservation of the particle-level collision/impact, a correction between numerical and physical duration of the impulses is proposed. Then, relaying on so-called time-scale correction of particle-level impulses (TPCI), which introduces a correction factor for the computation of the impulsive loads, new formulations for the source terms of the pressure Poisson equation (PPE) were derived based on density and divergence-free criteria. As a result,

TCPI-PND and TCPI-PND-DF source terms, directly dependents of the spatial discretization, are also proposed for more stable and robust computations.

In order to evaluate the performance of the proposed source terms, simulations were carried out considering 2D pure hydrostatic, steady pure hydrodynamic cases, and transient hydrodynamic impact case of dam breaking, as well as a 3D sloshing in a prismatic tank with cyclic and impulsive responses. The raw computed pressures obtained by the proposed source terms are compared to original ones and against analytical or experimental results.

The results show that the proposed source terms are more stable in time domain, with computed pressure nearly independent to time steps. As a result, beside increasing remarkably the robustness of the numerical method by extended the range of stable computation, they lead to significant improvement on the stability of the pressure computation by mitigating unphysical oscillations. Especially regarding the new TCPI-PND-DF source term, it achieved the most stable results with smooth pressure fields in all the situations tested in the presented study and is recommended for further applications. These features confirmed the effectiveness of the proposed time-scale correction of the particle-level impulses (TCPI) in the improvement of the stability of the computation.

From the viewpoint of practical engineering applications, instead of pressure relaxation coefficient that requires empirically calibration and adjustment, the only numerical parameter required in the proposed source terms is the propagation speed of the perturbations $c_s$, of which the calibration is much more straightforward due to its physical meaning. The calibration of the coefficient $c_s$ was conducted and a range of reference value was provided, see Eq. (50), based on hydrostatic cases. Also, in comparison to other spurious pressure oscillation mitigation strategies, the proposed approach is of extremely simple implementation and, the most relevant point, no additional computational effort is demanded to the already compute-intensive numerical method. Moreover, the proposed approach can be easily combined with the other strategies to further improve the stability and accuracy of pressure assessment.

Finally, the proposed source terms were applied and tested only for the MPS method. Nevertheless, they can be extended to other incompressible projection-based particle methods such as the ISPH or the consistent particle method (CPM) (Koh et al., 2012). Extension for multi-resolution MPS techniques (Chen et al., 2016; Tang et al., 2016; Shibata et al., 2017; Tanaka et al., 2018; Khayyer et al., 2019), in which high-resolution is used only near the local critical areas, while low-resolution is used in the far-field, should also be evaluated with respect to adaptivity and versatility of our code and is left for future studies.

## 8. Acknowledgments

This material is based upon research supported by the Office of Naval Research Global under the Award Number: N62909-16-1-2181. This study was financed in part by the Coordenação de Aperfeiçoamento de Pessoal de Nível Superior - Brasil (CAPES) - Finance Code 001. The authors are also grateful to Petrobras for financial support on the development of the MPS/TPN-USP simulation system based on MPS method and Frade Japão Petróleo Ltda for the financial support during the performance of the experimental sloshing tests. Finally, the authors are grateful to Dr. Pedro Cardozo de Mello for their valuable help in the execution of the experiments.

## 9. Appendix A

To illustrate the energy conservation properties of MPS, 2D standing wave simulations are being investigated by means of the comparisons between numerical and analytical solution for the free surface elevation at center of the tank obtained from second-order nonlinear wave theory (Wu & Taylor, 1994), and time histories of kinetic energy ($T$), potential energy ($U$) and total mechanical energy ($E_M$) computed as:

$$E_M = T + U \quad \text{with} \quad T = \frac{\rho l_0^{dim}}{2} \sum_{i \in \mathbb{F}} \|\mathbf{u}_i\|^2 \quad \text{and} \quad U = -\rho l_0^{dim} \sum_{i \in \mathbb{F}} \mathbf{g} \cdot \mathbf{r}_i \,. \tag{55}$$

where $\mathbb{F}$ represents the fluid particles.

Figure 31 shows the initial geometry of the 2D standing wave of mean water level $h_s = 1$m and length $\lambda_s = 2$m. The initial profile of water surface is given by:

$$\eta_0(x) = A_s \cos[k_1(x + \lambda_s/2)] \,, \tag{56}$$

where $\eta_0$ is the initial surface elevation (above the mean water level at 1.0m), $A_s = 0.1$m stands for the wave amplitude, and $k_1 = 2\pi/\lambda_s$ denotes the wave number. The fluid properties are density $\rho = 1000\text{kg/m}^3$ and kinematic viscosity $\nu_k = 0.0$(inviscid). The initial distance between particles $l_0 = 0.01\,\text{m}$, time step $\Delta t = 2 \times 10^{-4}\,\text{s}$, relaxation parameter $\gamma = 0.01$ and the propagation speed $c_s = 2\text{m/s}$ were adopted.

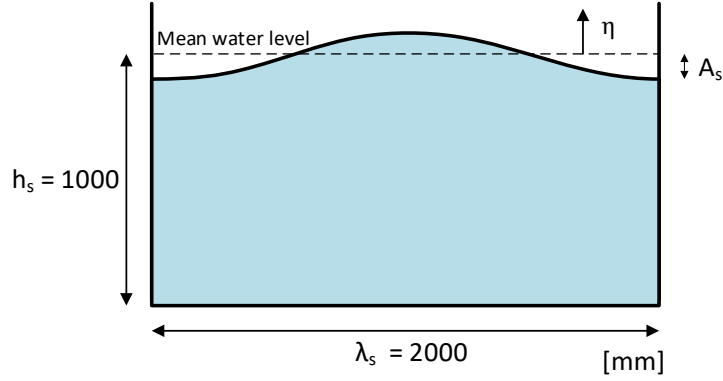

Figure 31: Schematic view of 2D standing wave. Geometry and main dimensions of the initial condition.

A 2$^{nd}$ order analytical solution by Wu and Taylor (1994) is adopted as the analytical solution given by:

$$\eta(t) = \eta_{1st}(t) + \eta_{2nd}(t), \tag{57}$$

$$\eta_{1st}(t) = A_s \cos(\omega_2 t), \tag{58}$$

$$\eta_{2nd}(t) = \frac{1}{8g}\left\{2(\omega_2 A_s)^2 \cos(2\omega_2 t) + \frac{A_s^2}{\omega_2^2}[k_2^2 g^2 + \omega_2^4 - (k_2^2 g^2 + 3\omega_2^4)\cos(\omega_4 t)]\right\}, \tag{59}$$

with

$$k_m = \frac{m\pi}{\lambda_s}; \quad \omega_m = \sqrt{k_m g \tanh(k_m h_s)}. \tag{60}$$

As previous reported by some authors (Khayyer et al., 2017; Wang et al., 2017), the energy conservation properties of particle-based methods are shown to be directly related with Taylor series consistency of pressure gradient models. In this way, simulations using original 0$^{th}$ order, see Eq. (9), and 1$^{st}$ order gradient correction, were performed herein. The conditionally 1$^{st}$ order-accurate corrected gradient model (CGM), i.e., 1$^{st}$ order consistency if the compact support is fully filled of a symmetric distribution of neighboring particles (Khayyer & Gotoh, 2013), proposed by Wang et al. (2017), is adopted for the pressure gradient:

$$\langle \nabla P \rangle_i = \left[\sum_{j \in \Omega_i} \omega_{ij} \frac{\mathbf{r}_{ij}}{|\mathbf{r}_{ij}|} \otimes \frac{\mathbf{r}_{ij}^T}{|\mathbf{r}_{ij}|}\right]^{-1} \sum_{j \in \Omega_i} \frac{P_j - \hat{P}_i}{\|\mathbf{r}_{ij}\|^2} \mathbf{r}_{ij} \omega_{ij}. \tag{61}$$

Figure 32 depicts the time histories of the water surface elevation at the center of the rectangular tank ($x = 1.0$m). All computed results using the original $0^{th}$ order pressure gradient, independent of the source term, present a considerable damping of wave height in relation to the analytical solution, as shown in Figure 32(a). Conversely, the numerical simulations using the CGM, see Figure 32(b), reproduce the wave heights in very good agreement with the $2^{nd}$ order analytical one, although slight damping still exists.

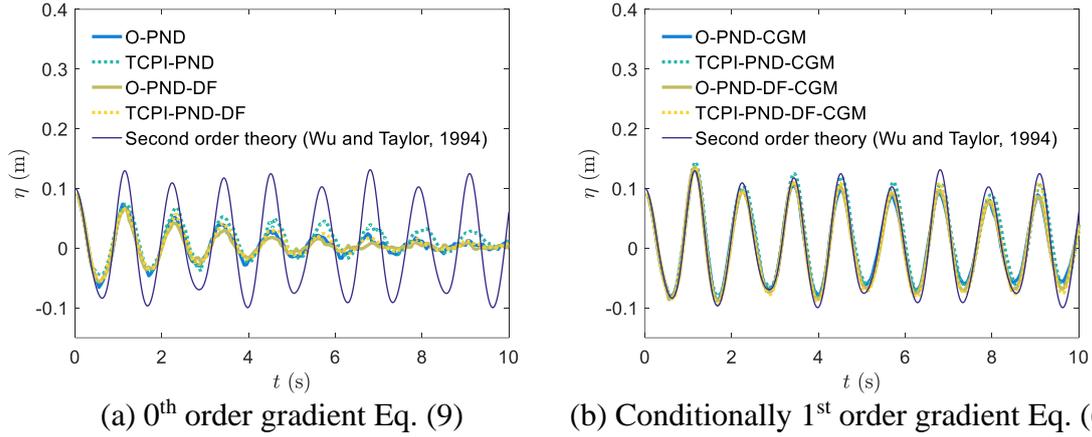

(a) $0^{th}$ order gradient Eq. (9)  (b) Conditionally $1^{st}$ order gradient Eq. (61)

Figure 32: 2D standing wave. Time histories of the water surface elevation at the center of the tank ($x = 1.0$m). Simulations using (a) $0^{th}$ order gradient given by Eq. (9) and (b) conditionally $1^{st}$ order-accurate corrected gradient model (CGM) given by Eq. (61), all adopting the particle distance $l_0 = 0.01$m and time step $\Delta t = 2 \times 10^{-4}$s.

Figure 33 shows the time histories of kinetic energy ($T$), potential energy ($U$) and variation of the mechanical energy ($E_M$) relative to the mechanical energy at the instant $t = 0$ ($E_{Mo}$). In Figure 33(a), the computed results using the original $0^{th}$ order pressure gradient show a considerable damping of the kinetic energy. On the other hand, the adoption of CGM leads to a notable improvement on the kinetic energy, as shown in Figure 33(b). After an initial stage of accommodation of the fluid particles, and, as consequence, sharp drop of potential energy for all simulations, see Figure 33(c) and (d), energy then rises rapidly because of surplus repulsive force provided by both $0^{th}$ order gradient and CGM, where the minimum pressure $\hat{P}_i$ is adopted, very similar to the results of standing waves obtained by Liu et al. (2018). Afterwards, the potential energy decreases due to the numerical dissipation. Moreover, Figure 33(d) shows that the computed potential energy evolutions using CGM are characterized by oscillations consistent with the expected standing wave motion. Conversely, the potential energy oscillation goes to zero with the time, i.e., notable wave damping, when the original $0^{th}$ order gradient is adopted, as depicted in Figure 33(c). The computed mechanical energy variation using original 0$^{th}$ order gradient and CGM are respectively within 1.5% and 1.2% of the initial mechanical energy, see Figure 33(e) and (f). In summary, these results indicate that high-order gradient, here the CGM, contributes significantly to the enhancement of the kinetic energy conservation and is highly recommended for long time simulations. Furthermore, it has been shown that the proposed source terms do not deteriorate the conservation features (volume and energy) of the projection-based MPS.

Kinetic energy

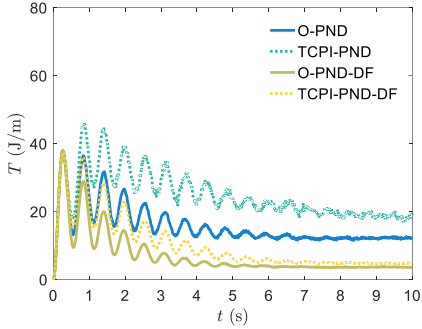

(a)

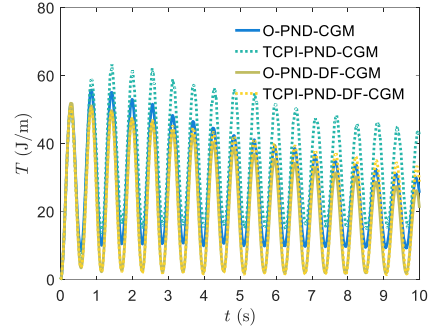

(b)

Potential energy

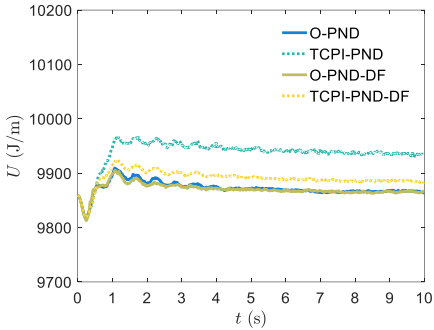

(c)

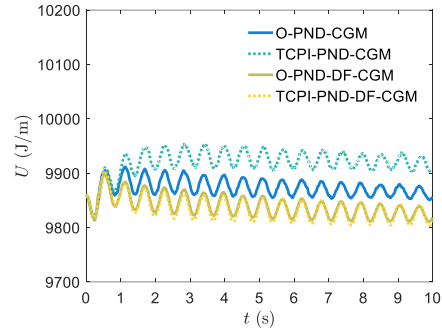

(d)

Mechanical energy variation

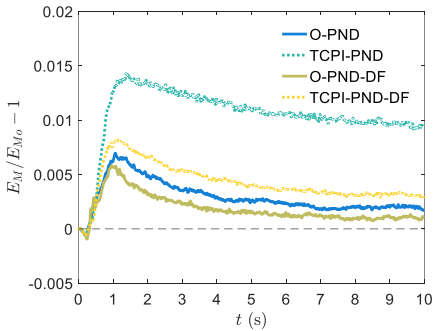

(e)

0$^{th}$ order gradient Eq. (9)

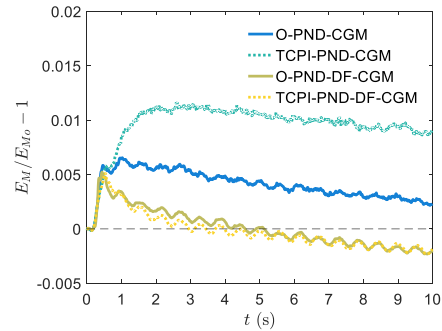

(f)

1$^{st}$ order gradient Eq. (61)

Figure 33: 2D standing wave. Evolution of energy components ($T$: Kinetic energy (top), $U$: Potential energy (middle), $E_M$: Mechanical energy and $E_{Mo}$: Mechanical energy at instant $t = 0$ (bottom)). Simulations using (left) 0$^{th}$ order gradient given by Eq. (9) and (right) conditionally 1$^{st}$ order-accurate corrected gradient model (CGM) given by Eq. (61), all adopting the particle distance $l_0 = 0.01$m and time step $\Delta t = 2 \times 10^{-4}$s.

Figure 34 show the numerical results at the instant $t = 7.94$s (7 periods), all obtained using the original $0^{th}$ order pressure gradient. The computed wave motion gradually decreased due to the kinetic energy damping and the free surface elevation is close to still water. The pressure fields computed by using O-PND and O-PND-DF are affected by a high-frequency oscillation resulting in unnatural pressure fields. Improvements on O-PND were achieved by adopting TCPI-PND, but undesirable noises remained, and the computed results using the TCPI-PND-DF show a smooth and accurate pressure field.

The better reproduction of wave shape, as consequence of less energy dissipation, is highlighted in Figure 35 for all simulations using CGM. The results using only the PND deviation, namely O-PND-CGM and TCPI-PND-CGM, show a significant level of pressure oscillation, while better and smooth pressure field are computed for both simulations using the PND deviation and divergence-free condition, namely O-PND-DF-CGM and TCPI-PND-DF-CGM.

In summary, the energy conservation is directly related to the order of precision of the discrete operator and the proposed source terms do not deteriorate the energy conservation features of the projection-based MPS. On the other hand, when the $1^{st}$ order operator is adopted, the contribution of the proposed source terms improved slightly the pressure computation based on PND. Moreover, for both PND and PND-DF, the proposed TCPI modification provided more robust computation, which is less sensitive to the time step.

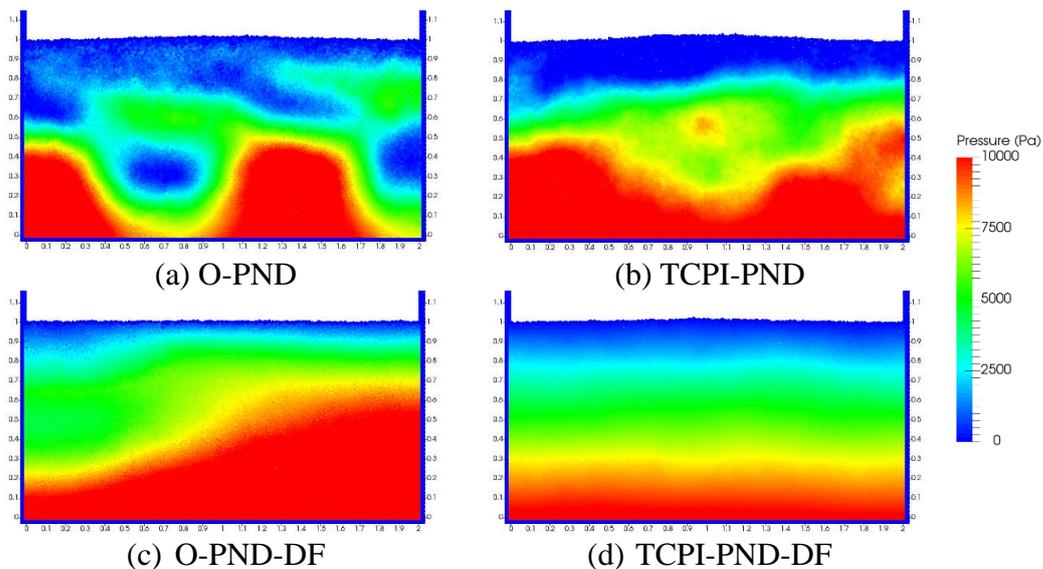

(a) O-PND  (b) TCPI-PND
(c) O-PND-DF  (d) TCPI-PND-DF

Figure 34: 2D standing wave pressure field at the instant $t = 7.94$s. Simulations using $0^{th}$ order gradient given by Eq. (9), all adopting the particle distance $l_0 = 0.01$m and time step $\Delta t = 2 \times 10^{-4}$s.

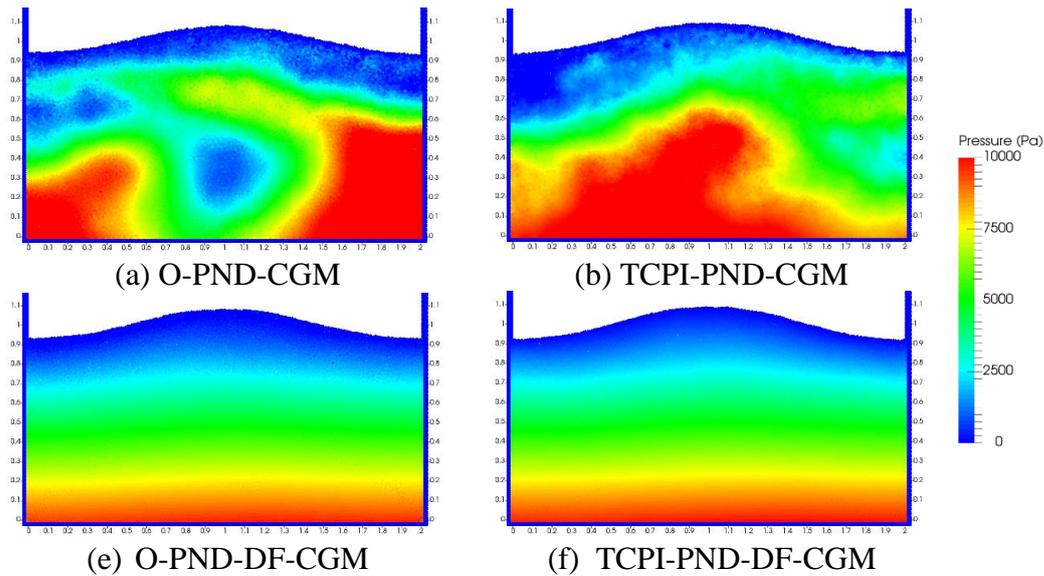

Figure 35: 2D standing wave pressure field at the instant $t$ = 7.94s. Simulations using conditionally 1st order-accurate corrected gradient model (CGM) given by Eq. (61), all adopting the particle distance $l_0 = 0.01$m and time step $\Delta t = 2 \times 10^{-4}$s.

Monaghan, J.J., 1992. Smoothed particle hydrodynamics. *Annual Review of Astronomy and Astrophysics*, 30, pp.543-74. Available at: https://doi.org/10.1146/annurev.aa.30.090192.002551.

Ng, K.C., Hwang, Y.H. & Sheu, T.W.H., 2014. On the accuracy assessment of Laplacian models in MPS. *Computer Physics Communications*, 185(10), pp.2412-26. Available at: https://doi.org/10.1016/j.cpc.2014.05.012.

Ngo-Cong, D., Tran, C.-D., Mai-Duy, N. & Tran-Cong, T., 2015. Incompressible smoothed particle hydrodynamics-moving IRBFN method for viscous flow problems. *Engineering Analysis with Boundary Elements*, 59, pp.172-86. Available at: https://doi.org/10.1016/j.enganabound.2015.06.006.

Ng, K.C., Sheu, T.W.H. & Hwang, Y.H., 2016. Unstructured moving particle pressure mesh (UMPPM) method for incompressible isothermal and non-isothermal flow computation. *Computer Methods in Applied Mechanics and Engineering*, 305, pp.703-38. Available at: https://doi.org/10.1016/j.cma.2016.03.015.

Pohle, F.V., 1951. *The Lagrangian equations of hydrodynamics: solutions which are analytic functions of the time*. New York: New York University.

Randles, P.W. & Libersky, L.D., 1996. Smoothed particle hydrodynamics: Some recent improvements and applications. *Computer Methods in Applied Mechanics and Engineering*, 139(1-4), pp.375-408. Available at: https://doi.org/10.1016/S0045-7825(96)01090-0.

Sandim, M., Paiva, A. & Figueiredo, L.H., 2020. Simple and reliable boundary detection for meshfree particle methods using interval analysis. *Journal of Computational Physics*, Available at: https://doi.org/10.1016/j.jcp.2020.109702.

Shakibaeinia, A. & Jin, Y.-C., 2010. A weakly compressible MPS method for modeling of open-boundary free-surface flow. *International journal for numerical methods in fluids*, 63(10), pp.1208-32. Available at: https://doi.org/10.1002/fld.2132.

Shibata, K., Koshizuka, S., Matsunaga, T. & Massaie, I., 2017. The overlapping particle technique for multi-resolution simulation of particle methods. *Computer Methods in*
66